\documentclass[a4paper,11pt]{article}
\usepackage[DIV12]{typearea}
\usepackage[latin1]{inputenc}
\usepackage[english]{babel}

\usepackage{cite}
\usepackage{amsfonts}
\usepackage{amsmath}    
\usepackage{amssymb}
\usepackage{cite}
\usepackage{mathrsfs}
\usepackage{pifont}
\usepackage{units}
\usepackage[small,loose]{subfigure}  
\usepackage{cancel}
\usepackage{ulem}
\usepackage[pdftex]{color}
\usepackage{slashed}
\usepackage{graphicx}

\usepackage[pdftex,colorlinks,pdfpagelabels]{hyperref}
\usepackage[figure,table]{hypcap} 
\hypersetup{
   bookmarksnumbered,
   pdfstartview={FitV},
   pdfpagemode={UseOutlines},
   pdfauthor={Rolf Kappl and Martin Wolfgang Winkler},
   pdftitle={},
   pdfsubject={scientific publication}
   ,citecolor={blue},
   linkcolor={blue},
   urlcolor={blue},
   filecolor={blue}
}

\newcommand{\eVdist}{\kern-0.06em}

\newcommand{\gev}{\:\text{Ge\eVdist V}}
\newcommand{\tev}{\:\text{Te\eVdist V}}

\newcommand{\pl}{p_{\text{\tiny{L}}}}
\newcommand{\pt}{p_{\text{\tiny{T}}}}

\newcommand{\f}[3]{\ensuremath{f_{\text{\tiny{$#1\!\!\rightarrow\!\! #2$}}}^{\text{\tiny{$#3$}}}}}
\newcommand{\n}[3]{\ensuremath{n_{\text{\tiny{$#1\!\!\rightarrow\!\! #2$}}}^{\text{\tiny{$#3$}}}}}

\newcommand{\iso}{\ensuremath{N_{\text{IS}}}}
\newcommand{\dsig}[4]{\ensuremath{\left(\frac{d\sigma}{#1}\right)_{\text{\tiny{$\!\!#2\!\!\rightarrow\!\! #3$}}}^{\text{\tiny{$\!\!#4$}}}}}
\newcommand{\dn}[4]{\ensuremath{\left(\frac{dn}{#1}\right)_{\text{\tiny{$\!\!#2\!\!\rightarrow\!\! #3$}}}^{\text{\tiny{$\!\!#4$}}}}}
\newcommand{\dnshort}[3]{\ensuremath{\left(\frac{dn}{#1}\right)_{\text{\tiny{$\!\!#2$}}}^{\text{\tiny{$\!\!#3$}}}}}
\newcommand{\dsigshort}[3]{\ensuremath{\left(\frac{d\sigma}{#1}\right)_{\text{\tiny{$\!\!#2$}}}^{\text{\tiny{$\!\!#3$}}}}}

\allowdisplaybreaks[1]

\begin{document}

\begin{titlepage}

\vspace*{-3.0cm}
\begin{flushright}
DESY 14-138
\end{flushright}

\begin{center}
{\Large\bf
  The Cosmic Ray Antiproton Background for AMS-02
}

\vspace{1cm}

\textbf{
Rolf Kappl$^a$,
Martin Wolfgang Winkler$^b$
}
\\[5mm]
\textit{$^a$\small
Bethe Center for Theoretical Physics \& Physikalisches Institut der 
Universit\"at Bonn, \\
Nu{\ss}allee 12, 53115 Bonn, Germany
}
\\[5mm]
\textit{$^b$\small
Deutsches Elektronen-Synchrotron DESY, \\
Notkestra{\ss}e 85, D-22607 Hamburg, Germany
}
\end{center}

\vspace{1cm}

\begin{abstract}
The AMS-02 experiment is measuring the cosmic ray antiproton flux with high precision. The interpretation of the upcoming data requires a thorough understanding of the secondary antiproton background. In this work, we employ newly available data of the NA49 experiment at CERN, in order to recalculate the antiproton source term arising from cosmic ray spallations on the interstellar matter. We systematically account for the production of antiprotons via hyperon decay and discuss the possible impact of isospin effects on antineutron production. A detailed comparison of our calculation with the existing literature as well as with Monte Carlo based evaluations of the antiproton source term is provided. Our most important result is an updated prediction for the secondary antiproton flux which includes a realistic assessment of the particle physics uncertainties at all energies.
\end{abstract}

\end{titlepage}

\tableofcontents

\vspace{2cm}

\section{Introduction}

Cosmic rays provide us with invaluable information about the most energetic processes happening in our universe. They also play an important role in the search for particle dark matter whose pair-annihilations might leave traces in the spectra of cosmic rays. In this light, the upcoming release of the antiproton measurement by the AMS-02 experiment~\cite{Aguilar:2013qda} at the international space station, is eagerly expected.

The dominant contribution to the antiproton flux in our galaxy -- the so-called secondary antiproton flux -- arises from the spallation of primary cosmic rays on the interstellar matter in the galactic disc. In order to identify a subdominant signal on top of the astrophysical background, the latter must be modeled precisely. This holds in particular as the antiproton signal from the most commonly considered dark matter candidates has a very similar spectral shape as the background flux.

The prediction of the secondary antiproton flux requires knowledge of the differential antiproton production cross section in proton proton and proton nucleus collisions. Existing parameterizations~\cite{Tan:1982nc,Duperray:2003bd} are mainly based on experimental data from the 1970s and early 80s. In this work, we revise those earlier parameterizations by making use of new precision data from the NA49 experiment at CERN~\cite{Anticic:2009wd,Baatar:2012fua}. Our approach also extends previous calculations in a number of ways:
\begin{itemize}
\item In a collider experiment -- depending on detector setup and kinematics -- antiprotons stemming from the decay of long-lived hyperons could partially escape detection. This would induce a systematic error in the measured antiproton production cross section. In this work we treat the antiprotons from hyperon decay separately and determine their production from the measured phase-space distribution of the parent hyperons.

\item A standard assumption in previous cosmic ray studies is the equal production of antiprotons and antineutrons in pp scattering. More recently, it was noted that isospin effects may induce an asymmetry in the number of $p\bar{n}$ and $n\bar{p}$ pairs leading to a preference for antineutron production~\cite{Fischer:2003xh,Chvala:2003dn}. We investigate the possible impact of isospin effects on the antiproton source term.
 
\item We improve the calculation of proton helium and helium helium scattering by making use of the simple empirical model introduced in~\cite{Baatar:2012fua}.

\end{itemize}
Our approach is to construct the total inclusive antiproton production cross section directly from the experimental data and well-established scaling arguments. For each step in our construction, we are able to include a realistic estimate of the related uncertainty. The resulting cross section is then translated into the secondary source term which we compare with previous calculations in the literature. As a further cross-check, we determine the source term with different Monte Carlo event generators.

Finally we use the two-zone diffusion model~\cite{Maurin:2001sj,Donato:2001ms} for describing the propagation of the antiprotons. As a result, we obtain a fully realistic prediction for the secondary antiproton flux, which includes a realistic assessment of the uncertainties in antiproton production down to lowest energies.

\section{Data-driven vs. Monte Carlo Approach}

The main uncertainty in the secondary antiproton source term is inherited from uncertainties in the inclusive cross sections which enter its determination. While dedicated attempts to evaluate the antiproton production cross sections using Monte Carlo generators have been performed~\cite{Simon:1998,Strong:1998pw,Donato:2001ms}, it is very difficult to estimate the uncertainties in the underlying hadronization models. In particular the low energy regime is a cause of concern as there exists no reliable theoretical description of soft hadronic processes. In this work, we therefore base our determination of the antiproton cross sections on experimental data. 
Nevertheless, several Monte Carlo tools will be used for the sake of comparison.

\subsection{Experimental Situation}

The dominant contribution to the antiproton source term arises from the scattering of cosmic ray protons with hydrogen in the galactic disc. One may think that in the LHC era proton proton scattering can be modeled with high precision. However, current collider experiments have been designed for the discovery of new physics. Detectors like CMS and ATLAS do not cover the high-rapidity region of the phase space, where most antiprotons are produced. Further, the energy scale relevant for cosmic ray experiments is considerably below the energy scale of operating colliders. AMS-02 is expected to detect antiprotons up to kinetic energies of several $100\gev$, which descend from primary cosmic rays with energies $E\simeq 10-10000\gev$. This corresponds to CMS energies $\sqrt{s}\simeq 5-100\gev$ of two colliding nucleons -- far below LHC energies.

The major part of proton proton scattering data in the relevant energy window were collected in the 1970s and early 80s. The most recent parameterization of the invariant antiproton cross section by Duperray et al.~\cite{Duperray:2003bd} fits these data with $\chi^2/\text{d.o.f.}\sim 4$~\cite{Duperray:2003bd}. While one would naively consider this a relatively poor fit, it is hardly possible to improve the level of consistency. The reason is that these data sets show a considerable scatter which complicates the determination of the differential cross section. Systematic errors in the old published data e.g. due to uncertainties in the beam luminosity are difficult to estimate. Certainly, a major source of error arises from the so-called ``feed-down problem'': About $1/4$ of the antiprotons in hadron collisions are produced via the decay of strange hyperons. Hyperons have a macroscopic decay length in the range cm to m, depending on their boost. One can only speculate which fractions of antiprotons from hyperon decay where properly identified by the old detectors. Clearly the fraction of escaping antiprotons depends on the boost of the parent hyperons and details of the detector setup. In particular, it varies between different regions of the phase space and between different experiments.

Fortunately, the situation has drastically improved due to the NA49 experiment at CERN. NA49 was a fixed target experiment with an incoming proton beam of energy $E_p=158\gev$ at the Super Proton Synchrotron. While the experiment was completed already in 2002, data on antiproton production were published only recently~\cite{Anticic:2009wd}. They provide a high phase-space coverage and drastically exceed the precision of previous data sets. The ``feed-down problem'' is absent as the antiprotons descending from hyperon decay have been identified using micro vertex detection and precision tracking. Therefore, we have decided to attempt a new determination of the antiproton source term based on the NA49 data. Different from previous approaches, we do not try to find a parameterization of the differential antiproton cross section, but we directly use the NA49 data as an input.

Apart from proton proton scattering, processes involving helium contribute considerably (in total about $40\%$) to the antiproton flux. Unfortunately, experimental data on $p$He scattering do not exist in the relevant energy window. Therefore, one has to extrapolate the helium cross sections from the measured cross sections of heavier nuclei. Fortunately, there exist new precision data on proton carbon scattering by NA49~\cite{Baatar:2012fua}. The collaboration has introduced a simple empirical model which relates the antiproton cross section in proton nucleus scattering with the cross section in proton proton scattering. This model was shown to describe $p$C scattering with good precision over the whole phase space and shall be used for $p$He and HeHe scattering in this work.

\subsection{Monte Carlo Generators}

As a possible alternative to the data-driven approach, Monte Carlo tools have been used to determine the secondary antiproton flux as well as a possible primary component from dark matter annihilations. In this light, it is important to investigate the range of applicability of the underlying hadronization models. In this work, we will employ the Monte Carlo generators PYTHIA 8.1~\cite{Sjostrand:2007gs}, DPMJET-III~\cite{Roesler:2000he} and GEANT4~\cite{Agostinelli:2002hh} for an independent determination of the differential antiproton production cross sections. For the simulation with PYTHIA we took into account all inelastic soft QCD processes. In the case of DPMJET, we used the implemented standard PHOJET model without elastic collisions. GEANT was developed as a detector simulation, but we adjusted the code to trace the event chain of single inelastic collisions. For the hadronization, we chose the build-in FTFP model which is based on the FRITIOF description of string fragmentation. The tool ROOT~\cite{Brun:1997pa} was used for data analysis and procession. 

PYTHIA only deals with proton proton interactions, therefore it can only be used to determine the dominant component of the antiproton source term. Subleading components from processes involving helium can be obtained by use of DPMJET and GEANT. As is pointed out in the documentation of the Monte Carlo generators, none of the three tools is suited for the low energy regime, where the hadronization models break down. At higher energies, however, reasonable agreement between the data-driven and Monte Carlo based evaluation of the antiproton source term is expected.

\section{Antiproton Production in Proton Proton Scattering}

Proton proton scattering is the dominant source of antiprotons in our galaxy. In hadronic collisions antiprotons are promptly produced due to the factorization of the colliding partons. Additionally, antiprotons descend as decay products of long-lived intermediate states like antineutrons or hyperons. Before we discuss the different contributions to the inclusive antiproton production cross section, we shall turn to the energy scaling of the cross section.

\subsection{Invariant Cross Section and Radial Scaling}

We are interested in the inclusive production of a hadron $h$ in the reaction $pp\rightarrow h + X$, where $X$ stands for the sum of the remaining final state particles. For this we introduce the Lorentz invariant cross section
\begin{equation}\label{eq:invcr}
 \f{pp}{h}{}= E_h \frac{d^3\sigma}{dp_h^3}= \frac{E_h}{\pi} \frac{d^2\sigma}{d\pl d\pt^2}\,,
\end{equation}
where $E_h$ is the energy of the detected hadron and $d^3\sigma/dp_h^3$ the differential cross section with respect to the three-momentum $\boldsymbol{p}_h$. The longitudinal and transverse components of $\boldsymbol{p}_h$ are denoted by $\pl$ and $\pt$ respectively. It is useful to express the invariant cross section in terms of $\pt$ and a scaling variable. The radial and Feynman scaling variables are defined as
\begin{equation}
 x_R = \frac{E^*}{E_{\text{max}}^*}\,, \qquad  x_f = \frac{\pl^*}{\sqrt{s}/2}\,,
\end{equation}
where $E^*$ and $\pl^*$ denote the energy and longitudinal momentum of $h$ in the center of mass frame. The maximal energy is determined as $E_{\text{max}}^*=(s-M_X^2+m_h^2)/(2\sqrt{s})$ with $M_X$ being the minimal mass of the recoiling particles $X$.

In~\cite{Carey:1974gf,Taylor:1975tm} a large set of experimental data was analyzed. It was shown that the invariant cross section approaches a radial scaling limit
\begin{equation}
 \f{pp}{h}{}(\sqrt{s},x_R,\pt) \;\longrightarrow\; \f{pp}{h}{}(x_R,\pt)\,
\end{equation}
for $\sqrt{s} \gtrsim 10\gev$ independent of the nature of the final state hadron. This is an enormous simplification as -- within the radial scaling regime -- the cross section at all center of mass energies can be deduced from the cross section at one energy. If, instead, one expresses the invariant cross section in terms of the scaling variable $x_f$, it also approaches a scaling limit~\cite{Feynman:1969ej}, however only at considerably higher energies~\cite{Taylor:1975tm}. Therefore, the use of $x_R$ as a scaling variable is clearly preferred.

From a theoretical viewpoint, constituent exchange models predict a power law behavior of the invariant cross section~\cite{Low:1975sv,Nussinov:1975mw,Brodsky:1976mg} 
\begin{equation}
\f{pp}{h}{}\propto (1-x)^n\,.
\end{equation}
Here $x$ denotes the light cone fraction of the considered hadron which is not directly observable, but can be approximated as $x\simeq x_R$~\cite{Brodsky:1977bu}. The fragmentation power $n$ is determined by the dimensional counting rule. It increases with the minimal number of spectator quarks necessitated by quantum number requirements. In particular, $n$ is expected to be larger for antibaryon production compared to meson production due to baryon number conservation. Quantitatively, the quark exchange model predicts the fragmentation power $n=9$ for antiproton production in $pp$ collisions~\cite{Brodsky:1977bu}.

\subsection{Contributions to the Inclusive Cross Section}

In cosmic ray physics, one is ultimately interested in the total number of antiprotons irrespective of their origin. However, in a collider experiment, antiprotons appearing macroscopically displaced from the initial collision vertex may escape detection. While antineutrons decay far outside any contemplable detector, the situation is more subtle for the strange hyperons $\bar{\Lambda}$ and $\bar{\Sigma}$ which have decay lengths comparable to typical detector scales. It was mentioned previously that this leads to the so-called ``feed-down problem'' as the fraction of escaping antiprotons from late decays remains dubious in the old experimental data. 

In this work, we mainly rely on antiproton data from the NA49 experiment. In contrast to previous experiments, antiprotons from hyperon decay have systematically been identified. This allows us to split the antiproton production cross section into individual components which we discuss separately. The total invariant antiproton cross section in pp collisions can be written as
\begin{equation}
f_{_{pp}}=  \f{pp}{\bar{p}}{}  +  \f{pp}{\bar{n}}{}\;.
\end{equation}
Here $\f{pp}{\bar{n}}{}$ accounts for the antiprotons from very late decaying antineutrons. This contribution is not directly accessible to experiments and has to be determined by use of symmetry arguments. We further split
\begin{equation}
\f{pp}{\bar{p}}{} = \f{pp}{\bar{p}}{0}+\f{pp}{\bar{p}}{\bar{\Lambda}}\;,
\end{equation}
where $\f{pp}{\bar{p}}{0}$ and $\f{pp}{\bar{p}}{\bar{\Lambda}}$ denote the contributions from prompt hadronization and from the weak decay of strange hyperons, respectively.

\subsubsection{Prompt Antiproton Production}\label{sec:prompt}

The NA49 collaboration has performed a precision measurement of inclusive antiproton production in $pp$ scattering~\cite{Anticic:2009wd}. Only the antiprotons from prompt hadronization are included in the data, while antiprotons from hyperon decay were systematically rejected. The beam energy $E_p=158\gev$ corresponds to $\sqrt{s}=17.4\gev$ -- well within the radial scaling regime. We can thus, in principle, use the NA49 data to determine the cross section at all energies $\sqrt{s}>10\gev$. 

\begin{figure}[t]
\begin{center}  
  \includegraphics[width=11cm]{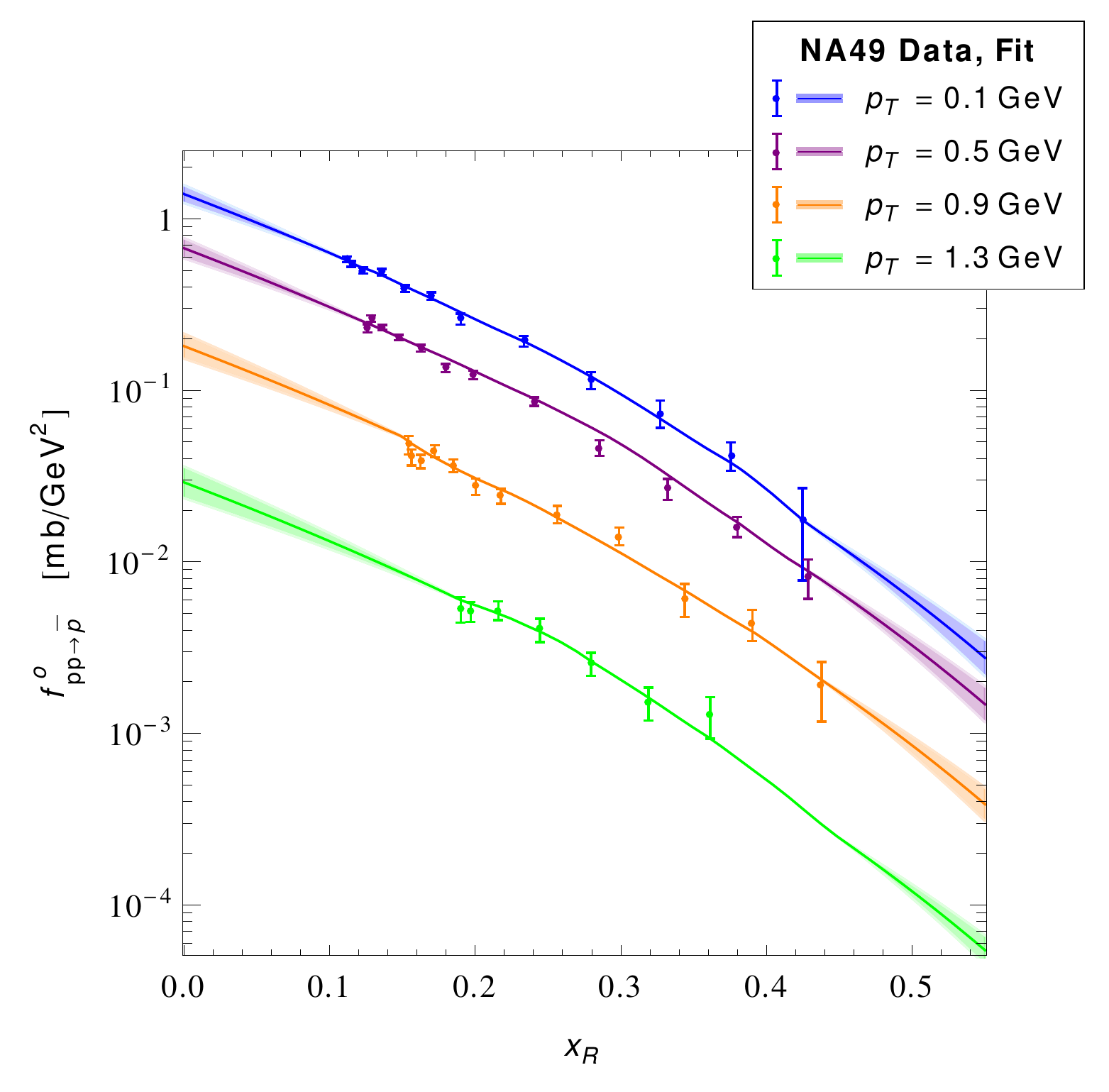}
\end{center}
\caption{Measured invariant antiproton cross section expressed in terms of the radial scaling variable and eyeball fit with extrapolation (see text).}
\label{fig:prompt}
\end{figure}

The collaboration provides the invariant differential cross section $\f{pp}{\bar{p}}{0}$ in bins of $x_f$ and $p_T$. Additionally, an eyeball fit to the data is specified. As Feynman scaling is only reached at considerably higher energies, it is convenient to translate $x_f$ into $x_R$. In figure~\ref{fig:prompt}, we depict a subset of the NA49 data and the corresponding eyeball fit in terms of the radial scaling variable.
Different from previous works, we do not attempt to find a parameterization for the invariant cross section, but simply use the fit given by the collaboration (after expressing it in terms of $x_R$). 

Despite the large phase-space coverage of NA49, we still have to use some extrapolation. The region at $x_R\simeq 0-0.1$ is kinematically not accessible at NA49 as $x_R > m_p/E_{\text{max}}^*$ by definition. Further, as the cross section decreases rapidly towards large $x_R$ and $p_T$, no data exist in this regime due to the limited statistics. In the covered region of phase space the invariant cross section shows virtually perfect power law behavior $\f{pp}{\bar{p}}{0} \propto (1-x_R)^n$ with $n=7.5$ at all $p_T$. We assume that this behavior continues outside the covered region of $x_R$ and include an uncertainty in the fragmentation power $n=7.5\pm 1$. This power law form is expected in theoretical hadronization models. As we will show explicitly later, our extrapolation is also consistent with experimental data taken at higher energies. We further observe an exponential decrease of the invariant cross section with the transverse mass $m_T=\sqrt{p_T^2+m_p^2}$ which we assume to continue at $p_T>1.5\gev$. We note, however, that this assumption is not of relevance as the cross section is negligibly small at high $p_T$.

Statistical errors in the NA49 data reside at the level of $\sim 5\%$. Their effect on the antiproton source term is, however, washed out by the phase space integration. They can be neglected compared to other sources of uncertainty. However, we shall take into account systematic errors which can affect the overall normalization of the cross section. Using a conservative linear error propagation, the NA49 collaboration has estimated the systematic uncertainty to be $6.5\%$ which we adopt in the following.

\subsubsection{Hyperons}\label{sec:hyperon}

The NA49 data set~\cite{Anticic:2009wd} includes only the promptly produced antiprotons, while antiprotons from hyperon decay were systematically rejected. In order to determine $\f{pp}{\bar{p}}{\bar{\Lambda}}$, we can use the measured phase-space distribution of the parent hyperons. In figure 22 of~\cite{Alt:2005zq}, the collaboration has published the differential multiplicity of $\bar{\Lambda}$-hyperons
\begin{equation}
 \dn{dx_{f_{\bar{\Lambda}}}}{pp}{\bar{\Lambda}}{}
 = \frac{1}{\sigma_{pp}^{\text{inel}}}\,\dsig{dx_{f_{\bar{\Lambda}}}}{pp}{\bar{\Lambda}}{}\;.
\end{equation}
where $\sigma_{pp}^{\text{inel}}$ denotes the total inelastic cross section in $pp$ collisions. The hyperons decay into antiproton and pion with a branching fraction $\text{Br}(\bar{\Lambda}\rightarrow \bar{p}\, \pi)=0.64$~\cite{Beringer:1900zz}. We can make use of the fact that $m_{\Lambda} \simeq m_p+ m_\pi$, which implies that the momentum of $\bar{\Lambda}$ is distributed between the decay products according to their masses. Therefore we can express the Feynman variable $x_f$ of the antiproton in terms of the Feynman variable of the parent hyperon
\begin{equation}
 x_f \simeq \frac{m_p}{m_\Lambda}\;x_{f_{\bar{\Lambda}}}\,.
\end{equation}
This allows us to predict the multiplicity distribution of antiprotons from hyperon decay 
\begin{equation}\label{eq:hyperontop}
\dn{dx_f}{pp}{\bar{p}}{\bar{\Lambda}}
 = \text{Br}(\bar{\Lambda}\rightarrow \bar{p}\, \pi)\times \frac{m_\Lambda}{m_p}\,
\dn{dx_{f_{\bar{\Lambda}}}}{pp}{\bar{\Lambda}}{}\;
+ (\bar{\Sigma} \text{ contribution})\;.
\end{equation}
Note that there is a subdominant contribution to $\n{pp}{\bar{p}}{\bar{\Lambda}}$ from \(\bar{\Sigma}\) hyperons. By symmetry arguments the ratio of produced $\bar{\Sigma}^-/\bar{\Lambda}$ can be estimated to be $0.33$~\cite{Anticic:2009wd}\footnote{According to~\cite{Anticic:2009wd}, the ratio of produced $\bar{\Sigma}^-/\Sigma^+$ is expected to be $0.8\,\bar{\Lambda}/ \Lambda$. If one uses the measured multiplicities of $\Lambda$ and $\Sigma^+$ from~\cite{Alt:2005zq} this leads to the estimate of $\bar{\Sigma}^-/\bar{\Lambda}\simeq 0.33$.}. 
We set the branching fraction $\text{Br}(\bar{\Sigma}^-\rightarrow \bar{p}\, \pi)=0.52$~\cite{Beringer:1900zz} and assume that the antiprotons from $\bar{\Sigma}$-decay follow the same $x_f$-distribution as the antiprotons from $\bar{\Lambda}$-decay.

\begin{figure}[t]
\begin{center}  
  \includegraphics[width=10.0cm]{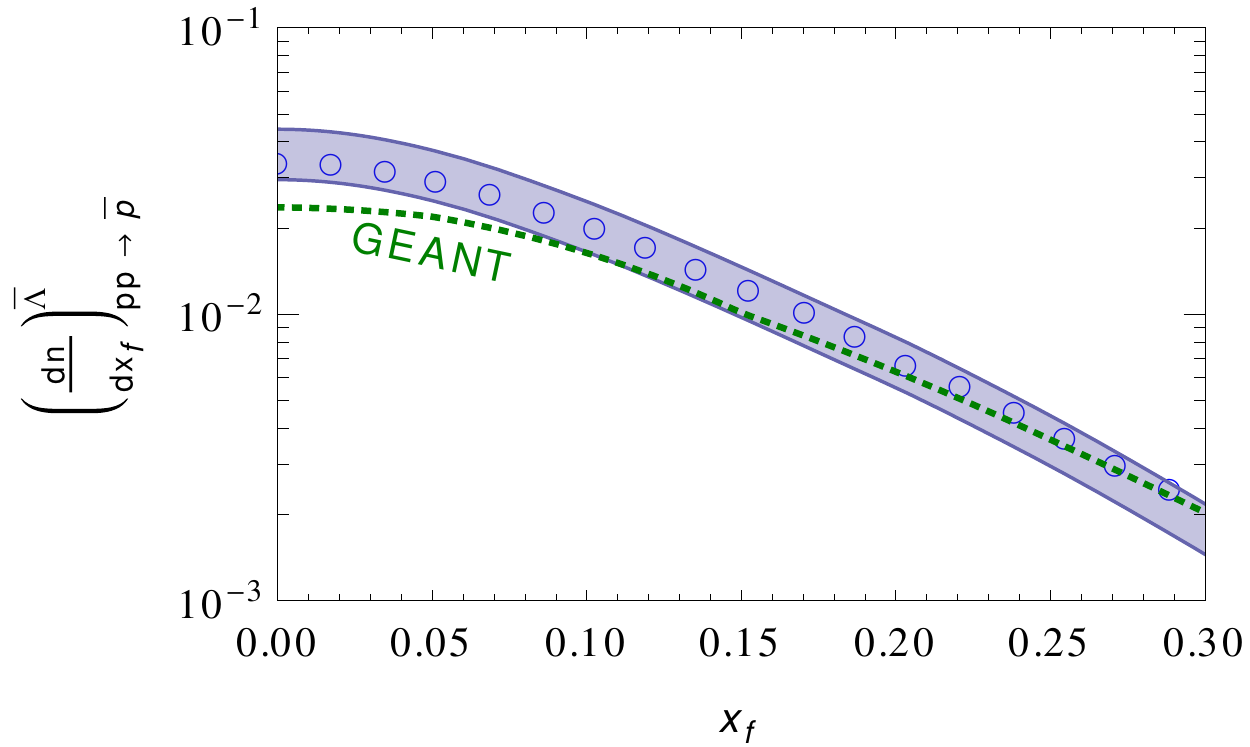}
\end{center}
\caption{Multiplicity distribution of hyperon-induced antiprotons integrated over transverse momenta as extracted from the NA49 data (blue circles). The prediction from the GEANT Monte Carlo is shown for comparison (green dashed line). The multiplicity distribution of promptly produced antiprotons scaled by $0.2 - 0.3$ is indicated by the blue band.}
\label{fig:hyperon}
\end{figure}

In figure~\ref{fig:hyperon}, we depict the differential multiplicity of antiprotons from hyperon decay. The data points were obtained from the measured differential multiplicity of $\bar{\Lambda}$~\cite{Alt:2005zq} by use of~\eqref{eq:hyperontop} (including the contribution from $\bar{\Sigma}$). For comparison, we also provide the distribution obtained with the GEANT Monte Carlo, which is in reasonable agreement with the data. The differential multiplicity of the promptly produced antiprotons scaled by a factor of $0.2$ and $0.3$ is also shown. The latter was determined by integrating the invariant cross section over transverse momenta
\begin{equation}\label{eq:multipli}
\dn{dx_f}{pp}{\bar{p}}{0}=\frac{\pi}{\sigma_{pp}^{\text{inel}}}\:\frac{\sqrt{s}}{2}\,\int dp_T^2\;\, \frac{\f{pp}{\bar{p}}{0}}{E_{\bar{p}}}\;.
\end{equation}
It can be seen that the antiprotons from hyperon decay and those from prompt hadronization exhibit a very similar momentum distribution. Unfortunately, error bars are not available for the hyperon data. But we presume that within the current precision, both distributions match up to a scaling factor. In the following we take the uncertainty band in $\f{pp}{\bar{p}}{\bar{\Lambda}}$ such that all data points of figure~\ref{fig:hyperon} are included. This leads to the estimate
\begin{equation}
 \frac{\f{pp}{\bar{p}}{\bar{\Lambda}}}{\f{pp}{\bar{p}}{0}} = 0.2-0.3\;.
\end{equation}
We do not require this ratio to be strictly momentum-independent. Rather, we assume that it only varies between $0.2$ and $0.3$ over the whole phase space.

\subsubsection{Antineutrons and Isospin Effects}\label{sec:isospin}

In previous cosmic ray studies, it was assumed that there is no distinction between $\bar{p}$ and $\bar{n}$ production in $pp$ collisions. Serious concerns about this hypothesis have been raised in~\cite{Fischer:2003xh}. There, data from proton collisions at different CMS energies were analyzed. It was noted that the net proton density, defined as the number of protons minus antiprotons per inelastic event, revealed a considerable energy dependence. At increasing $\sqrt{s}$, this quantity was found to grow dramatically towards low $x_f$. Baryon number conservation requires that baryons and antibaryons are produced in pairs. Under the assumption that the numbers of $\bar{p}n$ and $p\bar{n}$ pairs match, the above defined net proton density should correspond to the non-pair produced protons. The mentioned steep increase seems to indicate a problem with baryon number conservation -- given the assumption of equal numbers of $\bar{p}n$ and $p\bar{n}$ pairs holds. The authors of~\cite{Fischer:2003xh} conclude that there must be an asymmetry between $\bar{p}$ and $\bar{n}$ production.

While the inclusive process $pp\rightarrow \bar{n}+X$ is not directly observable, it is instructive to consider the flipped reaction $np\rightarrow \bar{p}+X$. We will decompose the antiproton multiplicity into a projectile and a target component
\begin{equation}
\dn{dx_f}{np}{\bar{p}}{}=\dn{dx_f}{np}{\bar{p}}{\text{\footnotesize{pro}}}\;+\;\dn{dx_f}{np}{\bar{p}}{\text{\footnotesize{tar}}}\,.
\end{equation}
The validity of this decomposition requires the independence of target and projectile factorization, such that the total multiplicity arises from the simple superposition of both contributions. This assumption has been experimentally verified in other hadronic interactions (e.g. by comparing $\pi p$ and $pp$ scattering~\cite{Barr:2006fs}).

In the forward ($x_f>0$) and backward ($x_f<0$) region, the multiplicity is dominated by projectile and target factorization respectively. However, there appears a small feed-over at $|x_f| \lesssim 0.1$, where both contributions slightly leak into the ``wrong'' hemisphere. One can define the target overlap function $F_\text{tar}(x_f)$ to project out the target contribution of the multiplicity
\begin{equation}
\dn{dx_f}{hp}{\bar{p}}{\text{\footnotesize{tar}}}=F_\text{tar}(x_f)\; \dn{dx_f}{pp}{\bar{p}}{}\,,
\end{equation}
where $h$ denotes an arbitrary baryon or meson projectile. The overlap function was found to be independent of the transverse momentum $p_T$ and of $\sqrt{s}$ if expressed in terms of $x_f$~\cite{Barr:2006fs}. We take $F_\text{tar}(x_f)$ from table~14 in~\cite{Baatar:2012fua}, the projectile overlap function is simply given as $F_\text{tar}(1-x_f)$ and fulfills the relation $F_\text{tar}(x_f)+F_\text{pro}(x_f)=1$.
\begin{figure}[t]
\begin{center}  
  \includegraphics[width=8.6cm]{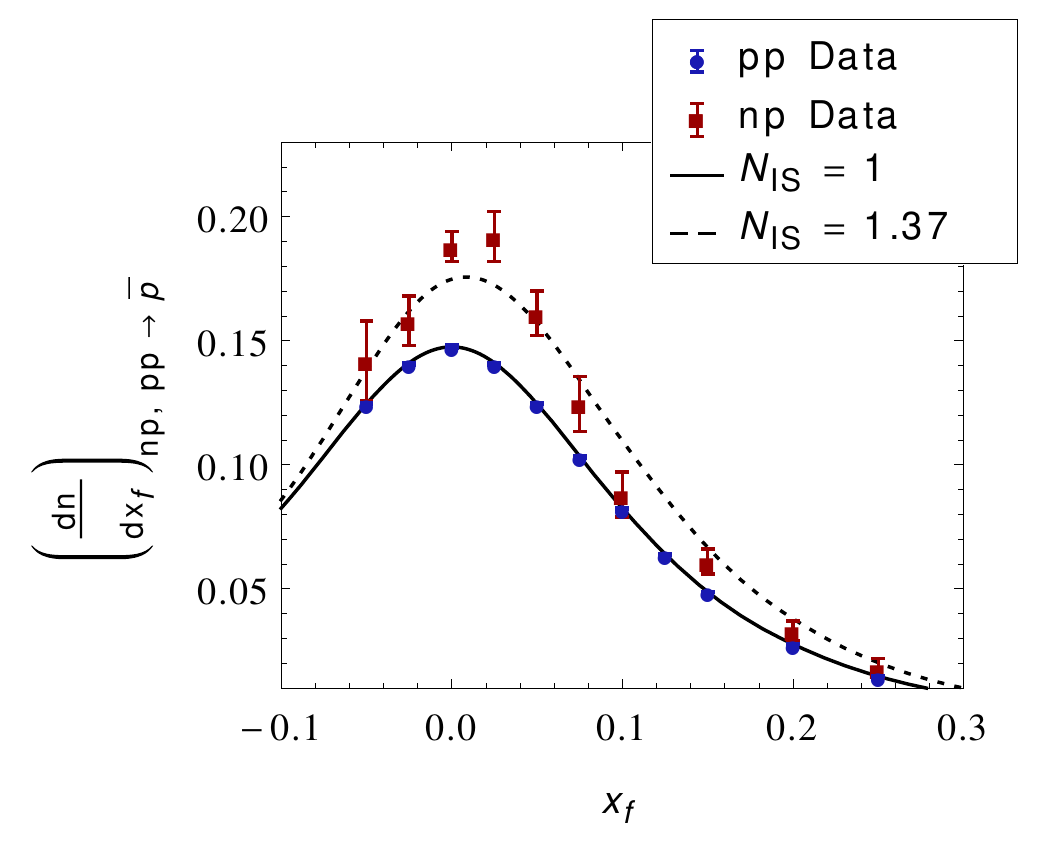}
\end{center}
\caption{Differential antiproton multiplicity in $pp$ and $np$ collisions. The difference between the two distributions can possibly be related to an isospin effect (see text).}
\label{fig:isospin}
\end{figure}
Using overlap functions to project out the target and projectile components, we can express
\begin{align}\label{eq:isomulti}
 \dn{dx_f}{np}{\bar{p}}{} &= F_\text{pro}(x_f)\: \dn{dx_f}{nn}{\bar{p}}{} +  F_\text{tar}(x_f)\:\dn{dx_f}{pp}{\bar{p}}{}\\
&=(\iso \;F_\text{pro}(x_f) +  F_\text{tar}(x_f))\, \dn{dx_f}{pp}{\bar{p}}{}\,.
\end{align}
In the second step, we made use of the fact that proton and neutron can be understood as doublet under an isospin symmetry which implies $\n{nn}{\bar{p}}{}=\n{pp}{\bar{n}}{}$. Further, we defined
\begin{equation}
\dn{dx_f}{pp}{\bar{n}}{}= \iso\;\dn{dx_f}{pp}{\bar{p}}{}\;. 
\end{equation}
Here $\iso$ denotes the isospin factor which parameterizes a possible asymmetry between $\bar{n}$ and $\bar{p}$ production in $pp$ collisions. 

In figure~\ref{fig:isospin}, we depict the antiproton multiplicity in $pp$ and $np$ collisions measured by NA49~\cite{Fischer:2003xh}. In the case $\iso=1$ both multiplicity distributions should match which appears inconsistent with the data. Indeed using a $\Delta\chi^2$-test and taking the isospin factor to be constant, we obtain $1.37\pm 0.06$ at the 90\% confidence level. The multiplicity distribution corresponding to the best fit point is also indicated in the figure.

This argumentation indicates a preference for the production of $p\bar{n}$ pairs compared to $n\bar{p}$ pairs in $pp$ collisions. One should note, however, that the $np$ data of NA49 are still at the preliminary level, systematic uncertainties have not been discussed. There are additional sources of concern: deviations from the independent target and projectile fragmentation would affect the above result. As we shall discuss later, the $p$C scattering data of NA49 are consistent with isospin effects, but hint at a somewhat smaller $\iso\sim1.2$. The hadronization models implemented in the Monte Carlo generators PYTHIA, GEANT and DPMJET yield no clear preference for $\bar{n}$ production at all. LHC data indicate that at very high energies, the $\bar{p}/p$ ratio in the mid-rapidity window approaches unity~\cite{Aaij:2012ut} which seems to speak against a preference for $p\bar{n}$ pairs in $pp$ collisions. Finally, it is not guaranteed that the isospin factor is constant in the whole phase space: the data in figure~\ref{fig:isospin} prefer a larger isospin effect at low $x_f$, although not at a statistically significant level.

As the current situation is inconclusive, a conservative treatment of the isospin effect seems appropriate. In the following, we will assume an isospin factor $\iso=1.0-1.43$. The lower end of this window corresponds to the standard assumption $\f{pp}{\bar{p}}{} = \f{pp}{\bar{n}}{}$, the upper end to the 95\% CL upper limit on the isospin factor deduced from $np$ scattering (see above). 

An additional contribution to antineutron production arises from hyperon decay, completely analogous as in the antiproton case (see section~\ref{sec:hyperon}). Assuming equal production of $\bar{\Sigma}^-$ and $\bar{\Sigma}^+$ and using the branching fractions from~\cite{Beringer:1900zz}, we find that the number of hyperon-induced antineutrons is by a factor of 1.05 higher than the number of hyperon-induced antiprotons.

\subsection{Corrections to Radial Scaling}

Let us now turn to the energy dependence of the invariant cross section. The latter is expected to be independent of the CMS energy for $\sqrt{s}>10\gev$ if expressed in terms of the radial scaling variable and the transverse momentum. This can be verified by considering high energy collider data. 

\begin{figure}[t]
\begin{center}  
  \includegraphics[width=10cm]{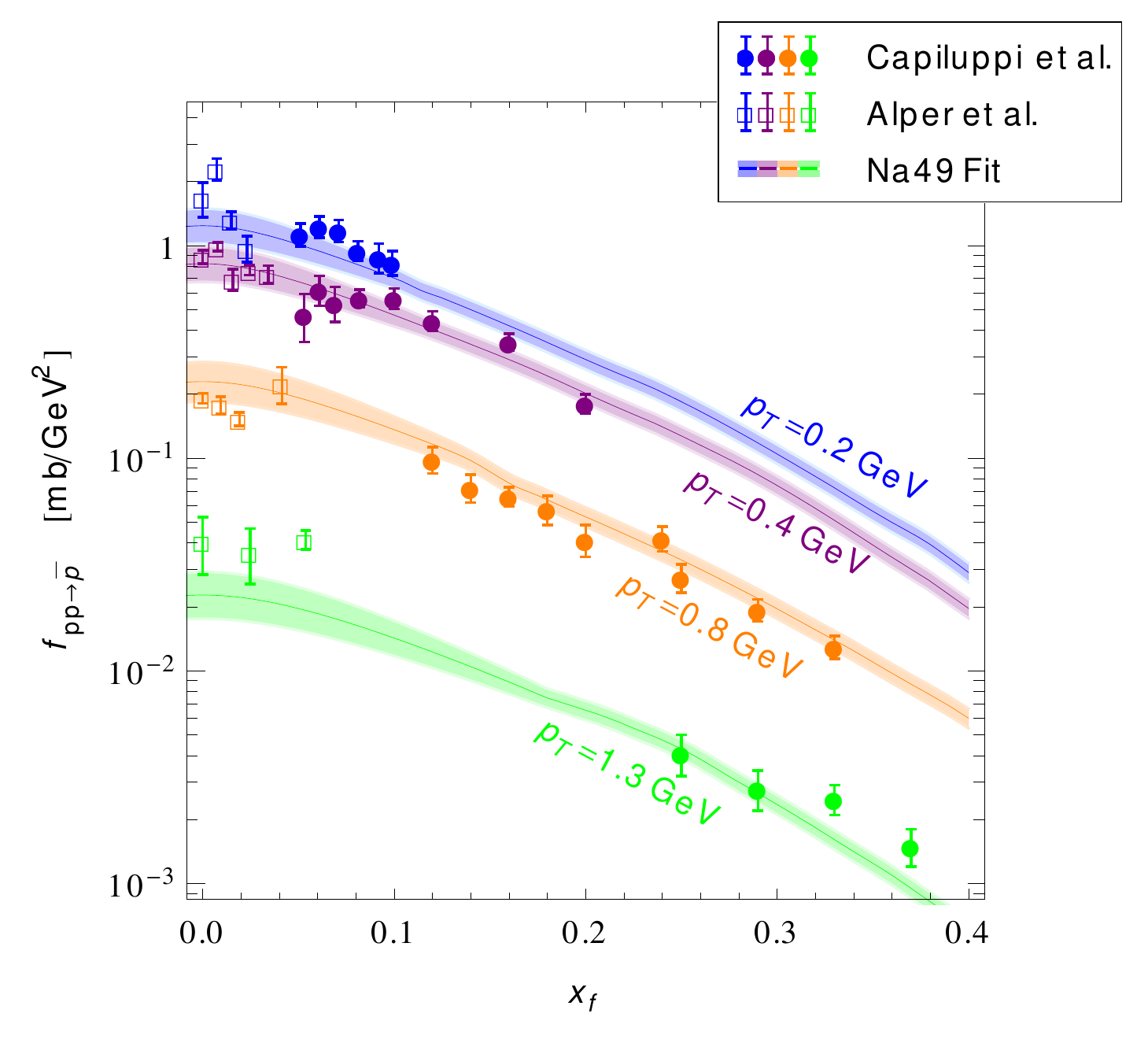}
\end{center}
\caption{Invariant antiproton cross section measured at CERN ISR compared with the fit used in this work (see text).}
\label{fig:radial}
\end{figure}

In figure~\ref{fig:radial}, we used the invariant cross section extracted from the NA49 data, to predict $\f{pp}{p}{}(x_f,p_T)$ at a higher CMS energy $\sqrt{s}=53\gev$.\footnote{Note that in the radial scaling regime $f(x_f,p_T)$ still depends on the CMS energy. This is because $x_f$ is a function of $x_R$ and $\sqrt{s}$. Only in the very high energy limit $x_f\rightarrow x_R$ and $f(x_f,p_T)$ becomes independent of $\sqrt{s}$. This corresponds to the Feynman scaling regime.} Corresponding experimental data from the CERN ISR collider by Capiluppi et al.~\cite{Capiluppi:1974rt} and Alper et al.\cite{Alper:1975jm} are also shown. We have included the antiprotons from hyperon decay in the NA49 prediction as a large fraction of them is expected to be contained in the CERN ISR data~\cite{Anticic:2009wd}. The indicated uncertainty band includes the systematic error of NA49, uncertainties related to our extrapolation of the NA49 data to low $x_R$ as well as the uncertainty in the hyperon contribution. Given that there is a considerable scatter in the CERN ISR data, one may take the shown experimental error bars with a grain of salt. But there is an overall good agreement between the measured cross section and the NA49 expectation.

\begin{figure}[t]
\begin{center}  
  \includegraphics[width=7.4cm]{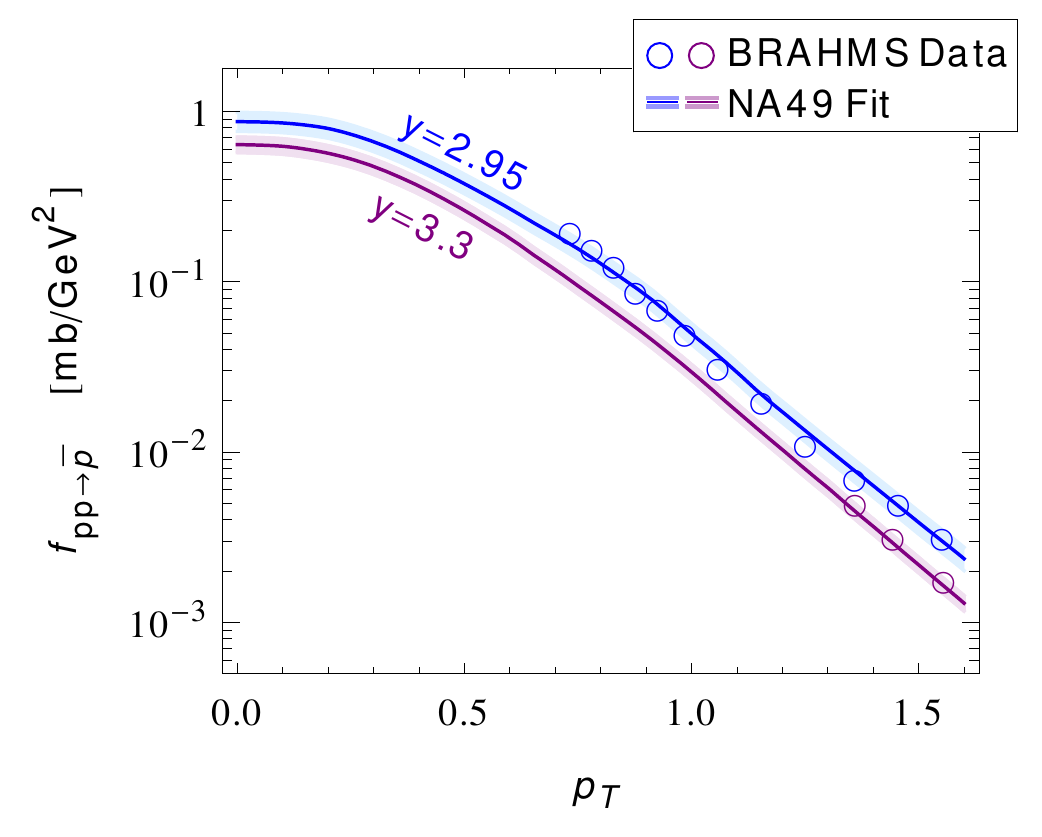}
\end{center}
\caption{Invariant antiproton cross section measured by the BRAHMS experiment compared with the fit used in this work (see text).}
\label{fig:fbrahms}
\end{figure}

For further comparison, we use a data set from the BRAHMS experiment at Brookhaven~\cite{Arsene:2007jd} which was taken at CMS energy $\sqrt{s}=200\gev$ for two different rapidities $y=2.95$ and $y=3.3$. The measured invariant cross section is in very good agreement with the NA49 prediction as can be seen in figure~\ref{fig:fbrahms}. This gives further confidence in the radial scaling hypothesis as well as our extrapolation of the NA49 data towards low $x_R$.

We note that deviations from radial scaling are expected at very high energies. The BRAHMS data indicate that, for $p_T<1.5\gev$, radial scaling is valid up to CMS energies $\sqrt{s}=200\gev$. Deviations from radial scaling at higher $p_T$ can be neglected as the cross section is anyway highly suppressed in this region of the phase-space. The BRAHMS energy corresponds to the scattering of an incoming proton with $E_p\sim 20\tev$ on the interstellar hydrogen. More energetic protons only marginally contribute to the antiproton flux in the energy window covered by AMS-02. Further, the growing uncertainty in the primary proton flux dilutes any effect caused by the violation of radial scaling at high $\sqrt{s}$. Therefore, we can safely work under the hypothesis of radial scaling for $\sqrt{s}>10\gev$.

At energies $\sqrt{s}\ll 10\gev$, however, radial scaling is heavily broken. In order to model the low energy antiproton flux, we clearly have to quantify the violation of radial scaling. We follow~\cite{Tan:1982nc} and define the energy-dependent ratio
\begin{equation}
 R(\sqrt{s},x_R,p_T) = \frac{f_{pp}(\sqrt{s},x_R,p_T)}{f_{pp}^{\text{RS}}(x_R,p_T)}\,,
\end{equation}
where $f_{pp}^{\text{RS}}(x_R,p_T)$ denotes the invariant cross section in the radial scaling limit. In~\cite{Taylor:1975tm} a large number of hadronic processes was considered. Independent of the inclusive channel, it was found that the radial scaling limit is approached asymptotically from above, i.e. $R(\sqrt{s},x_R,p_T) \geq 1$. Further, it was noted that the cross section enters the scaling regime faster at low $x_R$, while $R$ only weakly depends on $p_T$~\cite{Taylor:1975tm,Tan:1982nc}.

\begin{figure}[t]
\begin{center}  
  \includegraphics[width=10.5cm]{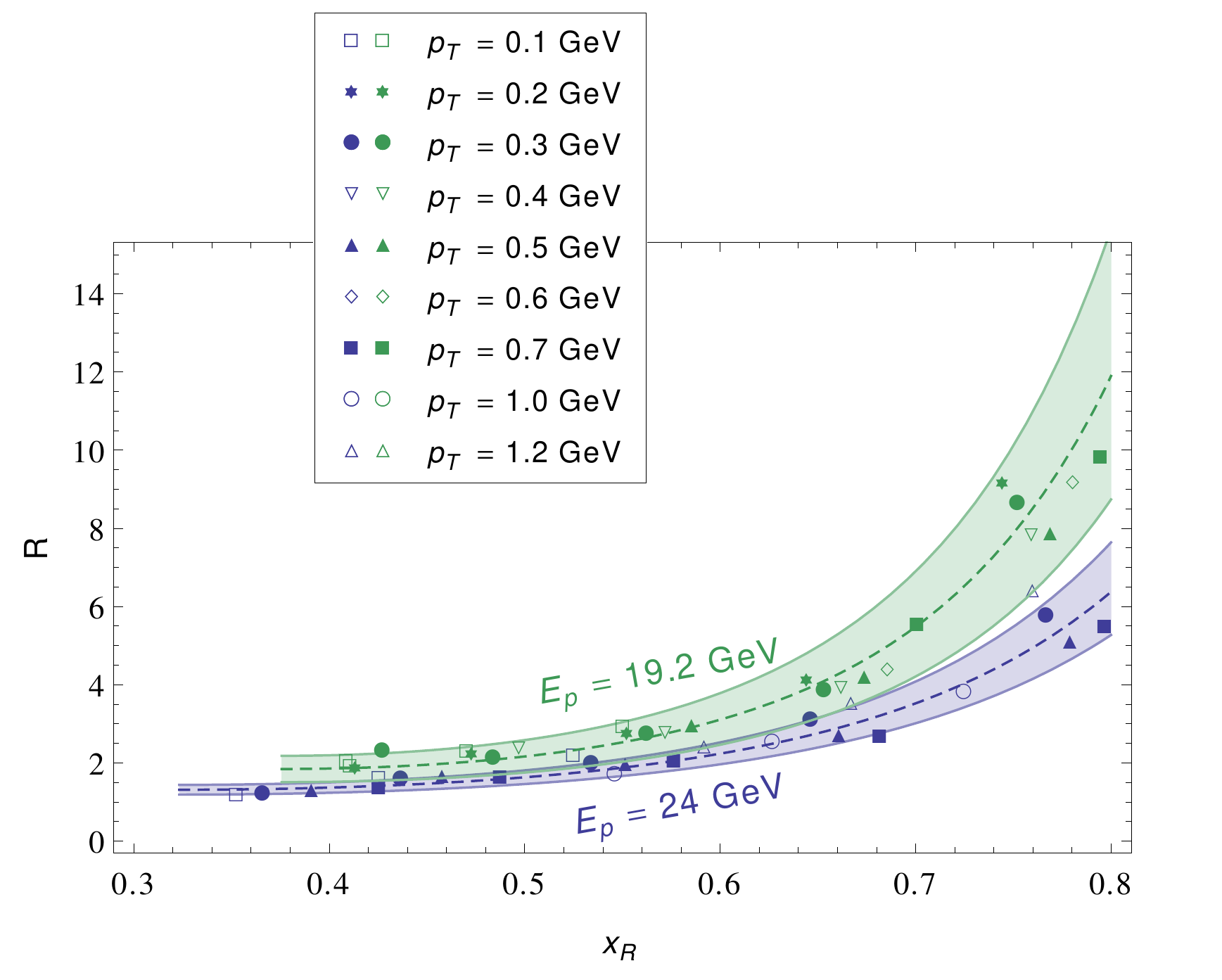}
\end{center}
\caption{Full invariant cross section divided by the cross section in the radial scaling regime at two different beam energies. Data points stem from the CERN Proton Synchrotron. Our parameterization of $R$ is indicated by the colored bands.}
\label{fig:lowe}
\end{figure}

Unfortunately, experimental data on inclusive $\bar{p}$-production at $\sqrt{s}<10\gev$ are rare. There exist only two data sets~\cite{1900hyj,Allaby:1970jt} with reasonable phase-space coverage. Both were recorded at the CERN Proton Synchrotron in the early 1970s. We have attempted to find a parameterization of $R$ with the qualitative behavior described above. For this, we neglected the dependence of $R$ on $p_T$, and required $R(x_R,\sqrt{s}=10\gev)=1$. Empirically we arrived at the function
\begin{equation}
 R(x_R,\sqrt{s}) = \left( 1 + C_1 \left(\frac{10\gev-\sqrt{s}}{\text{GeV}}\right)^5 \right)\exp \left[C_2 \,\left(\frac{10\gev-\sqrt{s}}{\text{GeV}}\right)\, (x_{\mathrm{R}}-x_{\mathrm{R},\text{min}})^2 \right]\,,
\end{equation}
for $\sqrt{s}\leq 10\gev$. The parameters were determined as $C_1=(1\pm 0.4)\times 10^{-3}$ and ${C_2= 0.7\pm 0.04}$. Given that error bars are only partially available for the data sets, we have chosen the uncertainties in $C_{1,2}$ such that the error band encloses the data points. Our parameterization as well as the experimental data~\cite{1900hyj,Allaby:1970jt} can be seen in figure~\ref{fig:lowe}. A slightly more involved function $R$ was suggested in~\cite{Tan:1982nc}. We have verified that the antiproton source term does virtually not depend on which of the two parameterizations we use. This is not surprising as~\cite{Tan:1982nc} employed the same limited amount of data available at low $\sqrt{s}$.

\subsection{Analytic Approximation}

In this work, we decided to use the fit to the Na49 data for the invariant antiproton cross section. To allow for a quick comparison with our results, we nevertheless want to specify an analytic approximation for $f_{_{pp}}$. In the NA49 data we observe an almost perfect factorization of $\f{pp}{\bar{p}}{0}$ with respect to the variables $x_R$ and $p_T$. A good fit ($\chi^2=1.4/\text{d.o.f.}$) can be obtained with the parameterization
\begin{equation}
 \f{pp}{\bar{p}}{0}(x_R,p_T) = \left(400\:\text{mb}\,\text{GeV}^{-2}\right) \times (1-x_R)^{7.76} \exp\left(-5.95\,m_T\right)\;,
\end{equation}
where $m_T$ stands for the transverse mass. To obtain $f_{_{pp}}$ one has to include the contributions from antineutron and hyperon decay. Further in the low energy regime the correction function $R$ must be applied. Using central values for the hyperon induced antinucleons, we arrive at
\begin{align}
 f_{_{pp}}(x_R,p_T,\sqrt{s})& \nonumber\\
= (400\:\text{mb} & \,\text{GeV}^{-2}) \times R(x_R,\sqrt{s})\times(1.51+\iso)\times (1-x_R)^{7.76} \exp\left(-5.95\,m_T\right)\;.
\end{align}
where we left the isospin factor unspecified. We have verified that the antiproton source term calculated from the analytic approximation agrees with the source term obtained from the NA49 fit within a few $\%$ precision.

\section{Antiproton Production in \textit{p}A and AA Scattering}

Given the lack of experimental data on $p$He and HeHe scattering, the corresponding differential cross sections must be determined by means of extrapolation. Duperray et al.~\cite{Duperray:2003bd} have therefore employed a modified version of the Kalinovskii formula~\cite{Kalinovskii:1989} which describes generic proton nucleus collisions. The parameters of the formula were fitted by using a large set of data, including scattering of protons on very heavy nuclei like lead. However, one should be aware that the nuclear medium effects in such scatterings depend crucially on the size of the nucleus. While they are almost absent for $p$He processes, they completely distort the phase space distribution of resulting hadrons in $p$Pb scattering. It is, therefore, questionable to which extent data involving heavy nuclei are useful for the extrapolation of the $p$He cross section. In this work, we construct the $p$He and HeHe cross sections from the elementary hadronic processes.

\subsection{Empirical Description}

\begin{figure}[t]
\begin{center}  
  \includegraphics[width=9cm]{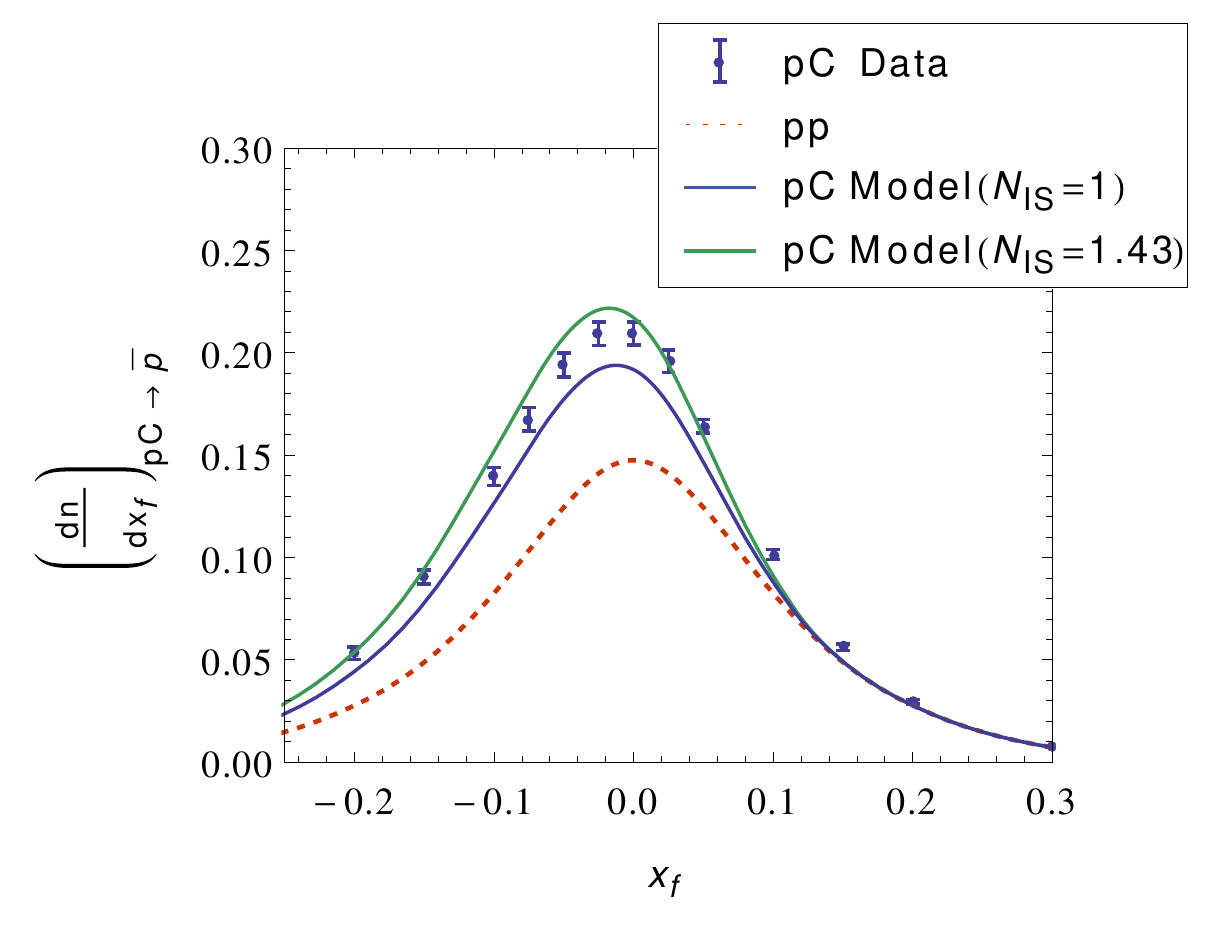}
\end{center}
\caption{Measured antiproton multiplicity in proton carbon scattering and prediction from the empirical model discussed in the text. The multiplicity in proton proton scattering is shown for comparison.}
\label{fig:carbon}
\end{figure}

In~\cite{Baatar:2012fua}, it has been realized that the differential antiproton multiplicity in $p$C scattering is closely related to the multiplicity in $pp$ scattering. Indeed, it was found that both multiplicities virtually match in the projectile hemisphere, i.e. in the forward direction. At the same time, a significant increase of the multiplicity is observed in the target hemisphere. The origin of this excess can be traced back to multiple scatters of the projectile in the nucleus. By measuring the increase of pion yields in the backward hemisphere, the average number of proton interaction was found to be $\langle \nu_{_C} \rangle=1.6$ in carbon. Using the overlap functions introduced in section~\ref{sec:isospin}, the multiplicity can be written as
\begin{equation}\label{eq:carbon}
\dn{dx_f}{pC}{\bar{p}}{}= \left( \frac{1+\iso}{2}\,\langle\nu_{_C}\rangle \, F_\text{tar}(x_f)+F_\text{pro}(x_f)\right) \dn{dx_f}{pp}{\bar{p}}{}\,.
\end{equation}
Note that in addition to multiple scatters, there appears a possible isospin enhancement due to the fact that the carbon nucleus contains half neutrons. In figure~\ref{fig:carbon}, we depict the measured antiproton multiplicity in $p$C collisions and the prediction derived from~\eqref{eq:carbon}. For the isospin factor we have used the uncertainty band $\iso=1-1.43$ as given in section~\ref{sec:isospin}. It can be seen that the empirical model yields a very good description of $p$C scattering within the given uncertainties. Minor deviations arising in the projectile hemisphere can be related to nuclear medium effects like the Cronin effect~\cite{Cronin:1973fd}. They reside, however, at the level of a few per cent for carbon and are expected to be completely negligible for helium.

\subsection[Cross Section for $p$He and HeHe Scattering]{Cross Section for $\boldsymbol{p}$He and HeHe Scattering}

In order to determine the $p$He, He$p$ and HeHe cross section we adopt the same method as for $p$C scattering. We assume that the differential antiproton multiplicity is given as
 \begin{equation}\label{eq:hemult}
\dnshort{dx_f}{A_1 A_2}{}
 = \left( \langle\nu_{_{A_1}}\rangle \, F_\text{tar}(x_f)+\langle\nu_{_{A_2}}\rangle \, F_\text{pro}(x_f)\right) \dnshort{dx_f}{pp}{}\,,\qquad (A_{1,2}=p,\text{He}) \;.
\end{equation}
Here $n_{_{A_1 A_2}}=\n{A_1 A_2}{\bar{p}}{}+\n{A_1 A_2}{\bar{n}}{}$ stands for the total antiproton multiplicity including the contributions from hyperon and antineutron decay. Note that all isospin effects are already contained in $n_{_{pp}}$, there appear no extra isospin factors. This is because isospin effects may induce an asymmetry between antiproton and antineutron production, but they do not affect the total number of antinucleons which are produced.

The average number of interactions is $\langle\nu_{_p}\rangle=1$ for protons. In the case of helium, we determine $\langle\nu_{_{\text{He}}}\rangle$ from the total inelastic cross section as
\begin{equation}
 \langle\nu_{_{\text{He}}}\rangle = 4\,\frac{\sigma_{pp\text{,inel}}}{\sigma_{p\text{He,inel}}}\,,
\end{equation}
where the factor 4 originates from the nucleon number of helium. Extracting the total inelastic proton helium cross section from~\cite{Letaw:1983}, we find $\langle\nu_{_{\text{He}}}\rangle=1.25$.\footnote{In~\cite{Letaw:1983} an interpolation formula for inelastic proton nucleus cross sections is presented. It is noted in the text that this formula overestimates the proton helium cross section by $20\%$. Applying this correction, one arrives at $\sigma_{p\text{He,inel}}=45\:\text{mb}\times 4^{0.7}/1.2$.} The invariant cross section is related to the multiplicity via~\eqref{eq:multipli}. Assuming the same transverse momentum distribution of antiprotons in proton and helium scattering, we arrive at the invariant cross section
\begin{equation}\label{eq:fnuc}
f_{_{A_1 A_2}} = \frac{\sigma_{A_1 A_2,\mathrm{inel}}}{\sigma_{pp,\mathrm{inel}}}\left( \langle\nu_{_{A_1}}\rangle \, F_\text{tar}(x_f)+\langle\nu_{_{A_2}}\rangle \, F_\text{pro}(x_f)\right)\:f_{_{pp}}\,,\qquad (A_{1,2}=p,\text{He}) \;.
\end{equation}
We will assume that~\eqref{eq:fnuc} holds independent of the CMS energy. Note that $f_{_{p\text{He}}}$ and $f_{_{\text{HeHe}}}$ inherit the uncertainties contained in $f_{_{pp}}$.

\section{The Secondary Antiproton Flux}

\subsection{The Source Term}

\begin{figure}[t]
\centering
\includegraphics[width=9.0cm]{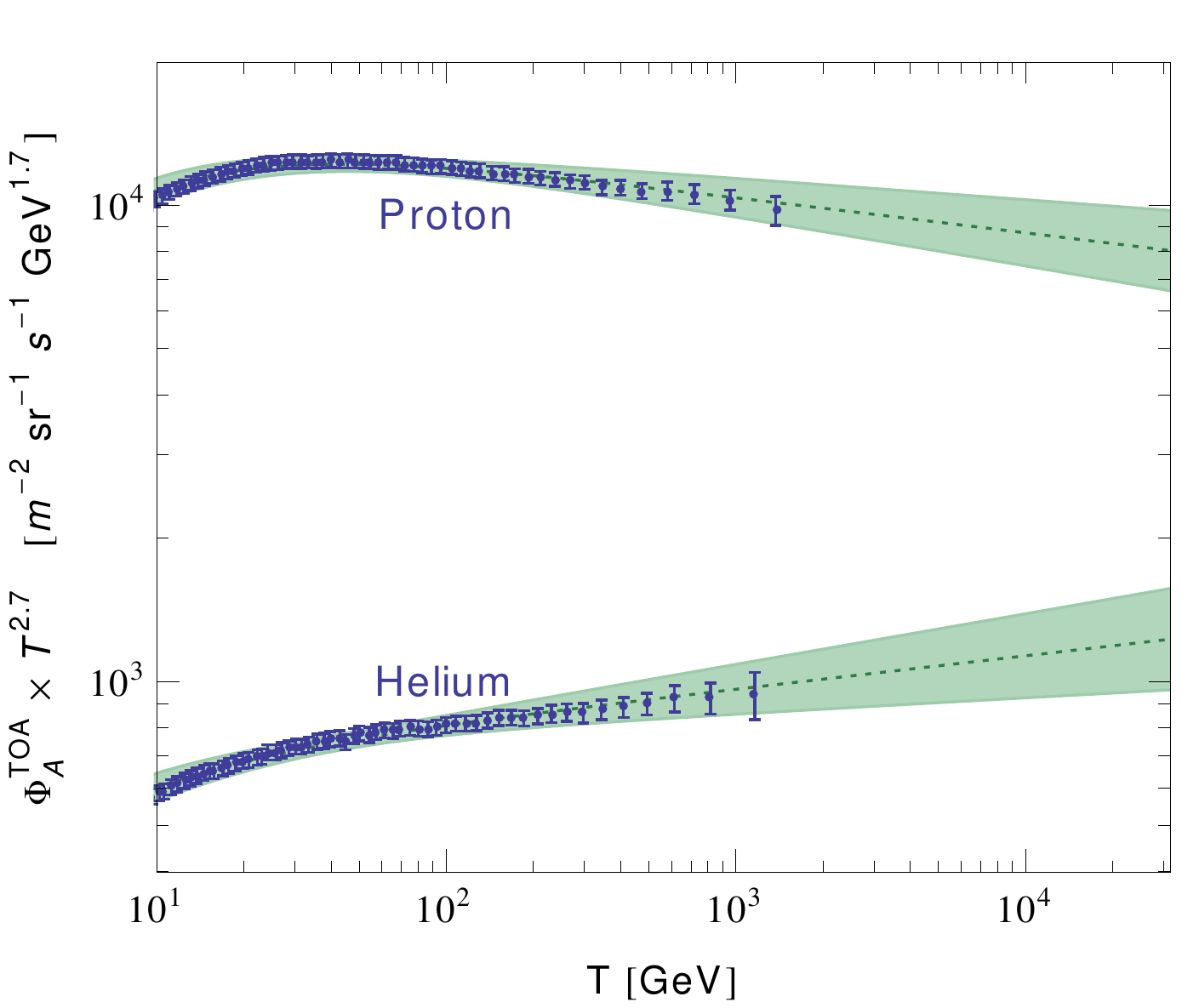}
\caption{Primary cosmic ray fluxes measured by AMS-02 and power law fits employed in this work (see text).}
\label{fig:jisfit}
\end{figure}

Primary cosmic rays create antiprotons by inelastic interactions with the interstellar gas in the galactic disc.
In the following, we want to determine the secondary antiproton source term $q_{\bar{p}}$ which is defined as the differential $\bar{p}$ production rate per volume, time and energy. For this, we make use of the invariant cross sections evaluated in the previous sections. To obtain the uncertainty band for the source term, we perform a conservative linear error propagation which is necessitated by the fact that errors are not normally distributed in our case.\footnote{In the case of isospin effects, there is e.g.\ a finite probability that $\iso=1$, which is at the edge of the considered uncertainty band.} We include the previously discussed uncertainties arising from systematic errors in the NA49 experiment, from the extrapolation of cross sections towards low $x_R$, from the hyperon-induced antinucleon production as well as from isospin effects. We also account for the uncertainty in the function $R$ which relates the invariant cross section at low collision energies with the cross section in the radial scaling regime. For our reference source term we use central values and set $\iso=1$.

The secondary source term is determined by
\begin{equation}
\label{eq:source}
 q_{\bar{p}}(T) = \sum\limits_{\text{A}_{1,2}=p,\text{He}}\;\,4 \pi\int\limits_{E_\text{th}}^{\infty} dT' 
\dsigshort{dT}{\text{A}_1\text{A}_2}{}
\varrho_{\text{A}_2} \;\;\Phi^{\text{IS}}_{\text{A}_1}(T')\;.
\end{equation}
Here $T'$ denotes the kinetic energy per nucleon of the incoming primary cosmic ray particle (proton or helium), $T$ the kinetic energy of the outgoing antiproton. The differential antiproton production cross section $(d\sigma/dT)_{_{\text{A}_1\text{A}_2}}$ can be obtained from the invariant cross section $f_{_{\text{A}_1\text{A}_2}}$ (cf.~\eqref{eq:invcr}). The threshold energy is given as $E_\text{th}=6\,m_p$, while for the interstellar number densities of hydrogen and helium, we set $\varrho_{p}=0.9\:\text{cm}^{-3}$ and $\varrho_{\text{He}}=0.1\:\text{cm}^{-3}$~\cite{Putze:2010zn}.\footnote{For a self-consistent approach, we have to employ the same input values for the hydrogen and helium densities as were used for the determination of the propagation parameters which we adopt later.} The contribution of heavier elements to $q_{\bar{p}}$ can be neglected~\cite{Simon:1998}.

The interstellar fluxes of primary cosmic rays $\Phi^{\text{IS}}_{p,\text{He}}$ exhibit a power law behavior in the energy range $T>10\gev$ relevant for the secondary antiproton production. They can in principle be taken from previous studies~\cite{Potgieter:2013cwj,Shikaze:2006je}. However, as a new precision measurement of proton and helium fluxes by AMS-02 became available~\cite{Ams:2013}, we decided to reevaluate the primary fluxes. Experimental data always refer to the top-of-the-atmosphere fluxes $\Phi^{\text{TOA}}_{p,\text{He}}$. Therefore, we have to account for the effects of solar modulation. This is done by means of the force field approximation~\cite{Gleeson:1968zza}. To obtain the interstellar fluxes, we have fitted power law functions to the AMS-02 data, which we modulated by a force field $\phi=700\:\text{MV}$.\footnote{The force field $\phi=700\:\text{MV}$ was determined by comparing the low energy proton flux measured by AMS-02 with the interstellar flux of~\cite{Potgieter:2013cwj}.}

\begin{figure}[t]
\centering
\includegraphics[height=6.9cm]{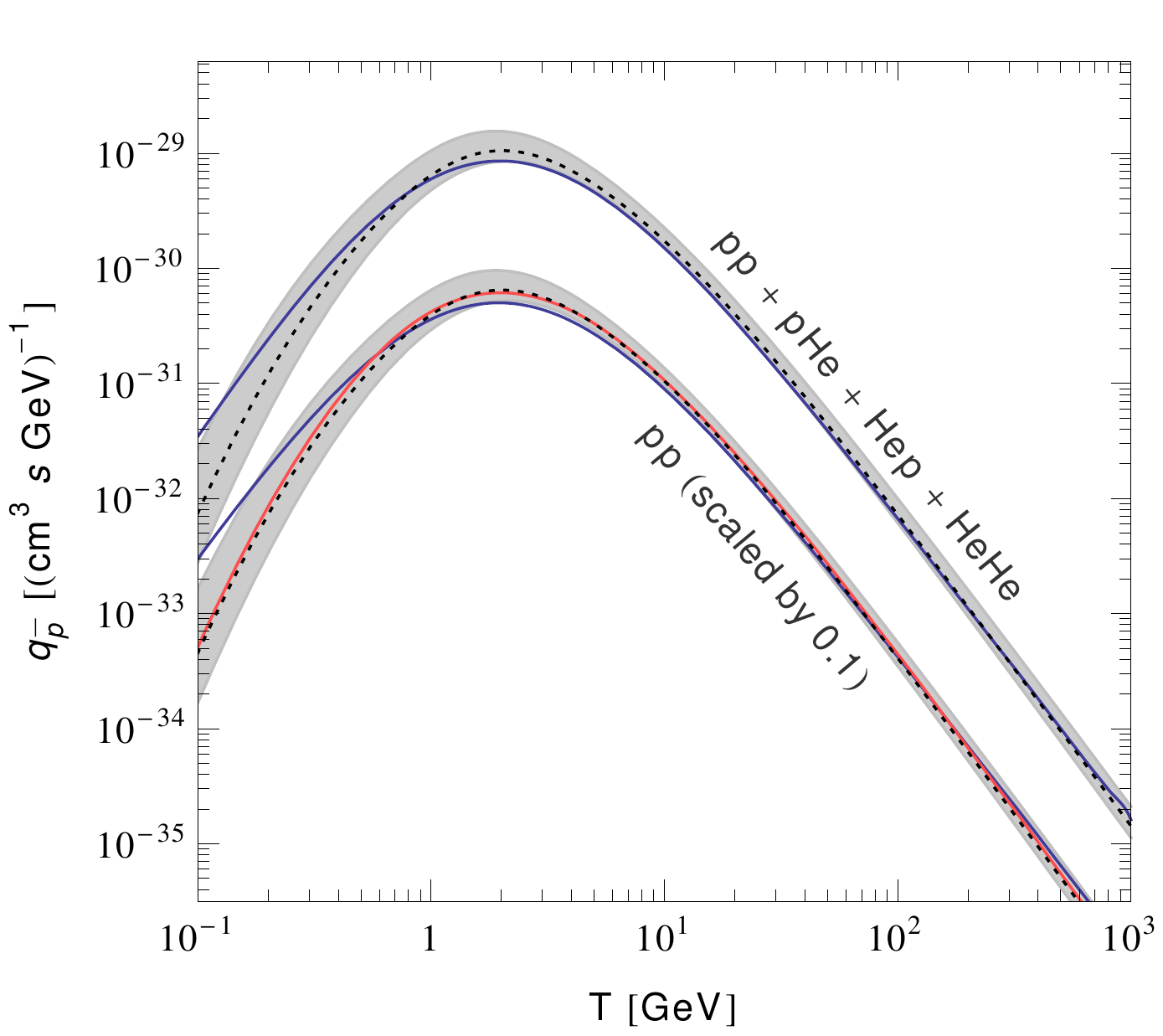}\hspace{-2mm}
\includegraphics[height=4.3cm]{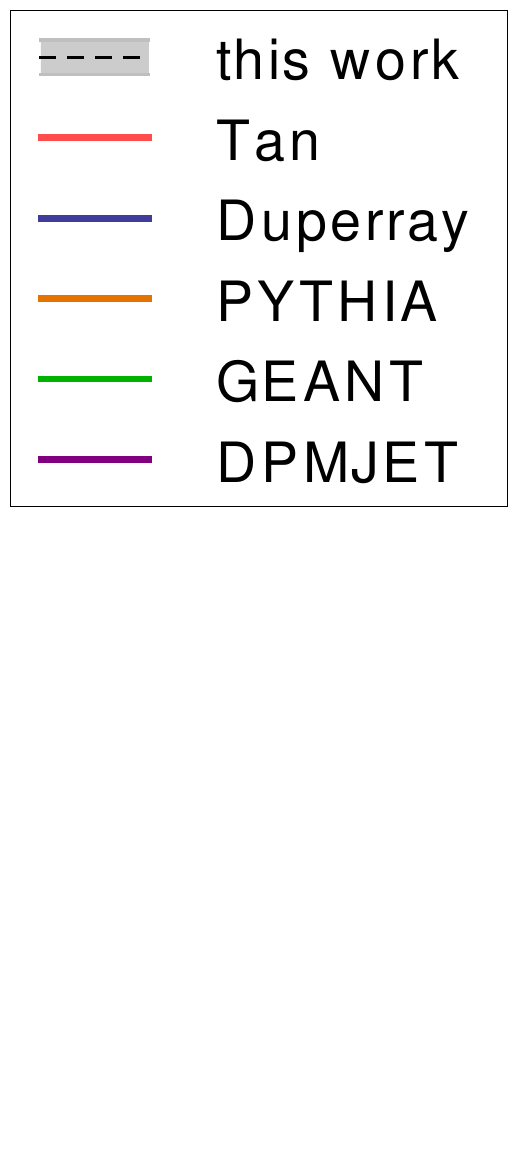}
\hspace{-6.5mm}
\includegraphics[height=6.9cm]{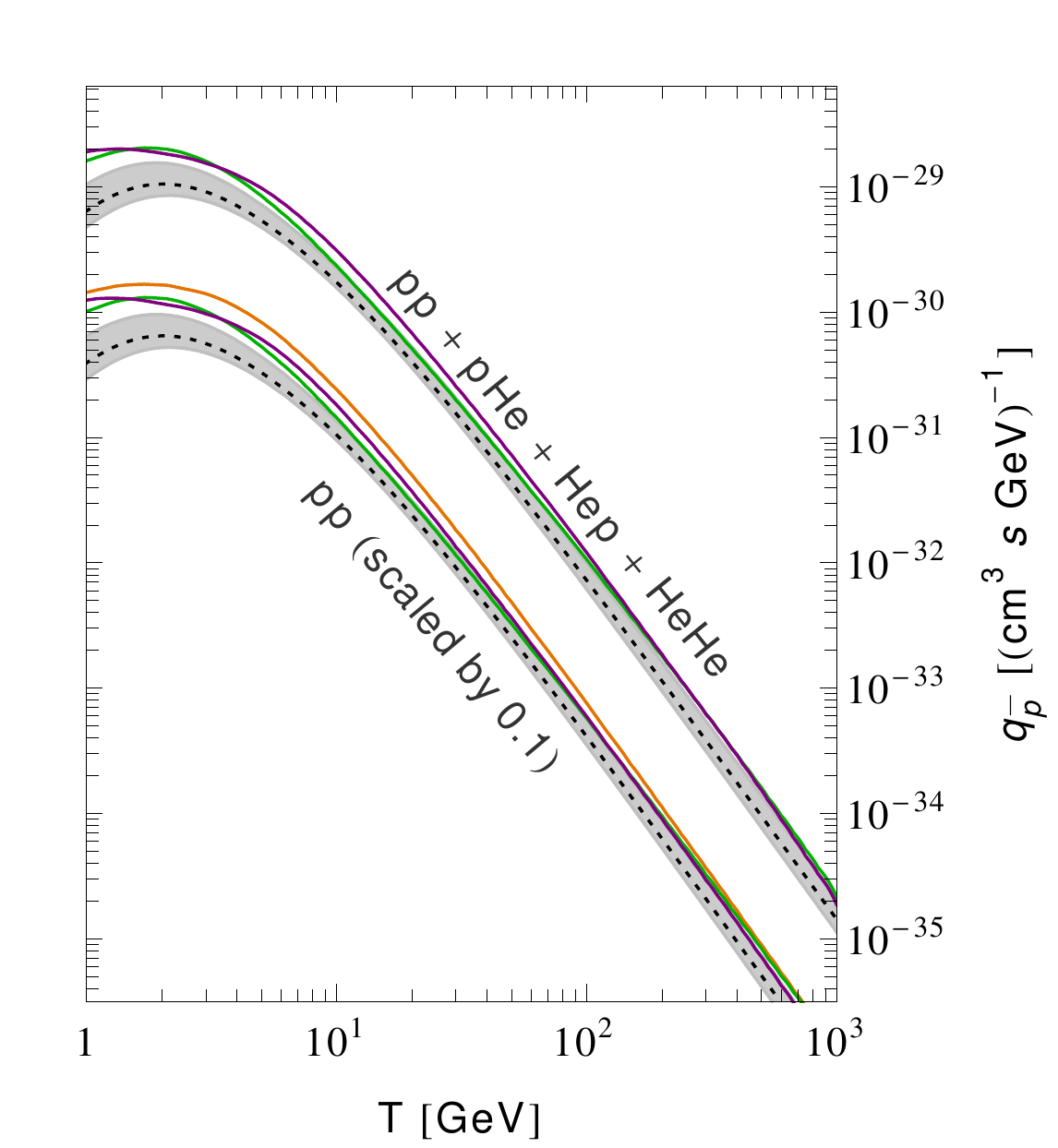}
\caption{Secondary source term determined from the NA49 data as described in the text and corresponding uncertainty band (also available in the Ancillary Files). The $pp$ component of the source term, scaled by 0.1, is shown separately. Also depicted are the source terms arising from previous parameterizations of the invariant cross sections (left panel). Source terms obtained with different Monte Carlo tools are shown in the right panel.}
\label{fig:na49vstan}
\end{figure}

In figure~\ref{fig:jisfit}, we depict the envelope of (modulated) power law functions which are consistent with the measured proton and helium fluxes (at the $90\%$ CL using a $\chi^2$-metric). In the calculation of the source term we include the shown bands as an additional source of uncertainty. For our reference source term we take the best-fit functions
\begin{align}
\Phi^{\text{IS}}_p(T)=17407\:\text{m}^{-2}\text{sr}^{-1}\text{s}^{-1}\text{GeV}^{-1}     \times \left(\frac{T}{\text{GeV}}\right)^{-2.775}\;,\\
\Phi^{\text{IS}}_\text{He}(T)
=597.2\:\text{m}^{-2}\text{sr}^{-1}\text{s}^{-1}\text{GeV}^{-1} \times \left(\frac{T}{\text{GeV}}\right)^{-2.630}\;.
\end{align} 
In figure~\ref{fig:na49vstan} we depict the antiproton source term which results from the invariant cross sections used in this work. The source term includes contributions from $pp$, $p$He, He$p$ and HeHe scattering, the $pp$ component is shown separately. The associated error in $q_{\bar{p}}$ resides at the level of $20\%$ for intermediate energies $T\sim 10\gev$. In this regime it is dominated by the uncertainty in the isospin factor which affects antineutron production. Towards low $T$, the width of the error band increases drastically and reaches $\sim 50\%$ at $T=1\gev$. This results from the uncertainty in the hadronic cross sections due to the breakdown of radial scaling and the lack of low-energy collider data. Towards high kinetic energies, the error in $q_{\bar{p}}$ also rises which is related to the growing uncertainties in the primary cosmic ray fluxes.

For comparison we have determined $q_{\bar{p}}$ by using previous parameterizations of the invariant antiproton cross section by Tan et al.~\cite{Tan:1982nc} and Duperray et al.~\cite{Duperray:2003bd} (see left panel of figure~\ref{fig:na49vstan}). In the $pp$ channel the Tan source term virtually matches with our reference source term. This is remarkable as our description of $pp$ scattering relied on a disjoined set of experimental data compared to Tan et al.. Duperray et al. have attempted to find a single parameterization of the invariant cross section $f_{_{pp}}$ in- and outside the radial scaling regime. This induces a tendency to underestimate $f_{_{pp}}$ at high energies and to overestimate it at low energies. We believe that this explains the shape of the Duperray source term. We note, however, that the latter still resides within the uncertainty band of our $q_{\bar{p}}$.

In the right panel of figure~\ref{fig:na49vstan}, we show $q_{\bar{p}}$ as obtained with the Monte Carlo tools PYTHIA, GEANT and DPMJET. While the Monte Carlo tools qualitatively reproduce the expected shape of the source term, they generically predict too high $q_{\bar{p}}$. At $T>10\gev$ the GEANT and DPMJET source terms are only marginally outside the uncertainty band of our data-based source term, while PYTHIA already overestimates $q_{\bar{p}}$ by a factor of 2. The level of discrepancy increases continuously towards low $T$ which is most likely caused by the hadronization models, arriving at the edge of their validity. 
We have traced back the discrepancy between data-based and Monte Carlo approach to the fact, that the Monte Carlo generators predict too large multiplicities already in $pp$ collisions. As an example, the Monte Carlos find $\n{pp}{\bar{p}}{}=0.006,\, 0.005,\, 0.008$ (GEANT, DPMJET, PYTHIA) at a collision energy $\sqrt{s}=6.1\gev$ which has to be compared with the experimental value $\n{pp}{\bar{p}}{}=0.002$~\cite{Antinucci:1972ib}. Therefore, it seems that the tested Monte Carlo tools can not be used ``out of the box'' to determine the secondary antiproton flux.

\subsection{Propagation in the Diffusion Model}

The antiproton flux induced by the source term $q_{\bar{p}}$ is determined through the diffusion equation 
\begin{equation}
\label{eq:diffusionequation}
 \nabla (-K \:\nabla N_{\bar{p}} + \boldsymbol{V}_c \,N_{\bar{p}}) + 
\partial_T (b_\text{tot} \,N_{\bar{p}} -K_{EE} \:\partial_T N_{\bar{p}} ) 
+ \Gamma_\text{ann}\,N_{\bar{p}} = q_{\bar{p}}+ q_{\bar{p}}^{\text{ter}}\;, 
\end{equation} 
where $N_{\bar{p}}$ denotes the antiproton space-energy density, \(K\) accounts for diffusion and \(\boldsymbol{V}_c\) is the galactic wind. The function \(b_\text{tot}\) describes energy losses, ionization and reacceleration, it can be taken from~\cite{Maurin:2002ua}, \(K_{EE}\) is the so-called energy diffusion coefficient~\cite{Maurin:2002hw} and \(\Gamma_\text{ann}\) the annihilation rate which describes the antiproton annihilation with interstellar matter. The tertiary source term $q_{\bar{p}}^{\text{ter}}$ arises from inelastic scattering of secondary antiprotons in the galactic disc.

We solve this equation semi-analytically within the two-zone diffusion framework introduced in~\cite{Maurin:2001sj,Donato:2001ms}. Further details can be found in our previous publication~\cite{Kappl:2011jw}.\footnote{For a full numerical approach to the diffusion equation, see~\cite{Strong:1998pw,Moskalenko:2001ya,Evoli:2008dv}.} In the two-zone diffusion model, propagation depends on five parameters: the height of the diffusion cylinder\footnote{See~\cite{Lavalle:2014kca,Bringmann:2011py,Lavalle:2010yw} for recent discussions on $L$.} $L$, the galactic wind $V_c$, the Alfv\'en speed $V_a$ which enters $K_{EE}$ as well as the diffusion parameters $K_0$ and $\delta$ which are related to \(K\) as 
\begin{equation}
 K=K_0\beta\left(\frac{p}{\gev}\right)^\delta\;.
\end{equation}
Here $\beta$ and $p$ stand for the antiproton velocity and momentum. We use the propagation parameters shown in table~\ref{tab:proppara} which were taken from a recent boron to carbon analysis~\cite{Putze:2010zn}.
\begin{table}[ht]
\centering
\begin{tabular}{cccccc}
\(\delta\)&\(K_0\ (\text{kpc}^2\cdot\text{Myr}^{-1})\)
&\(L\ (\text{kpc})\)&\(V_c\ (\text{km}\cdot\text{s}^{-1})\)&\(V_a\ 
(\text{km}\cdot\text{s}^{-1})\)\\
\hline
0.86&0.0042&4&18.7&35.5
\end{tabular}
\caption{Propagation parameters taken from the boron to carbon analysis in~\cite{Putze:2010zn}.}
\label{tab:proppara}
\end{table}
As the focus of this study is on the particle physics uncertainties contained in the antiproton flux, we do not attempt to include the uncertainties in the propagation parameters\footnote{See e.g.~\cite{Evoli:2011id} for a recent discussion of astrophysical uncertainties.}. The latter can be obtained by propagating the antiproton source term with all configurations consistent with the boron to carbon ratio and taking the envelope of the obtained antiproton fluxes (see e.g.~\cite{Bringmann:2006im}).

\subsection{Prediction for the Antiproton Flux and Comparison with Data}

\begin{figure}[t!]
\centering
\includegraphics[width=11.5cm]{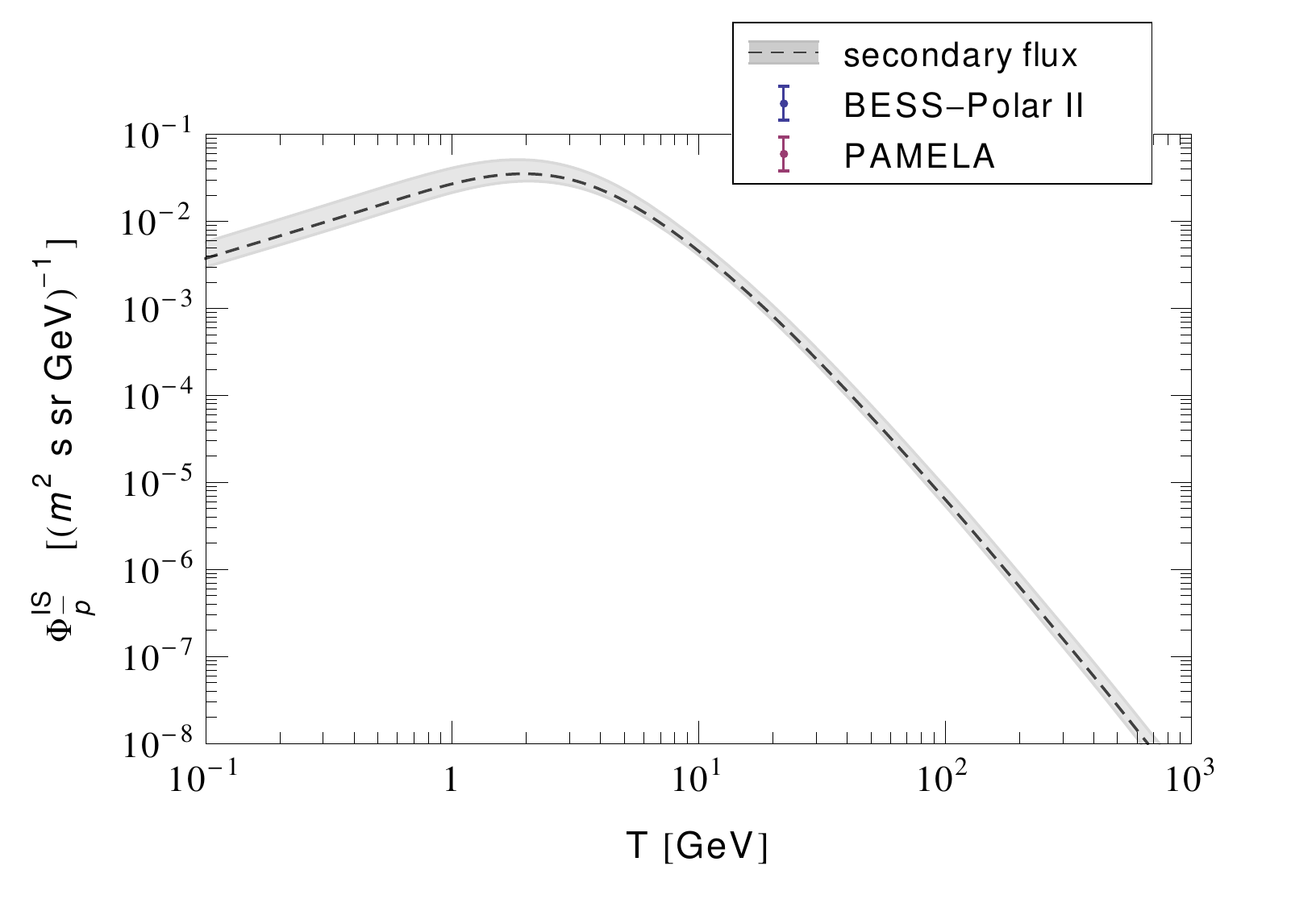}\\
\includegraphics[width=11.5cm]{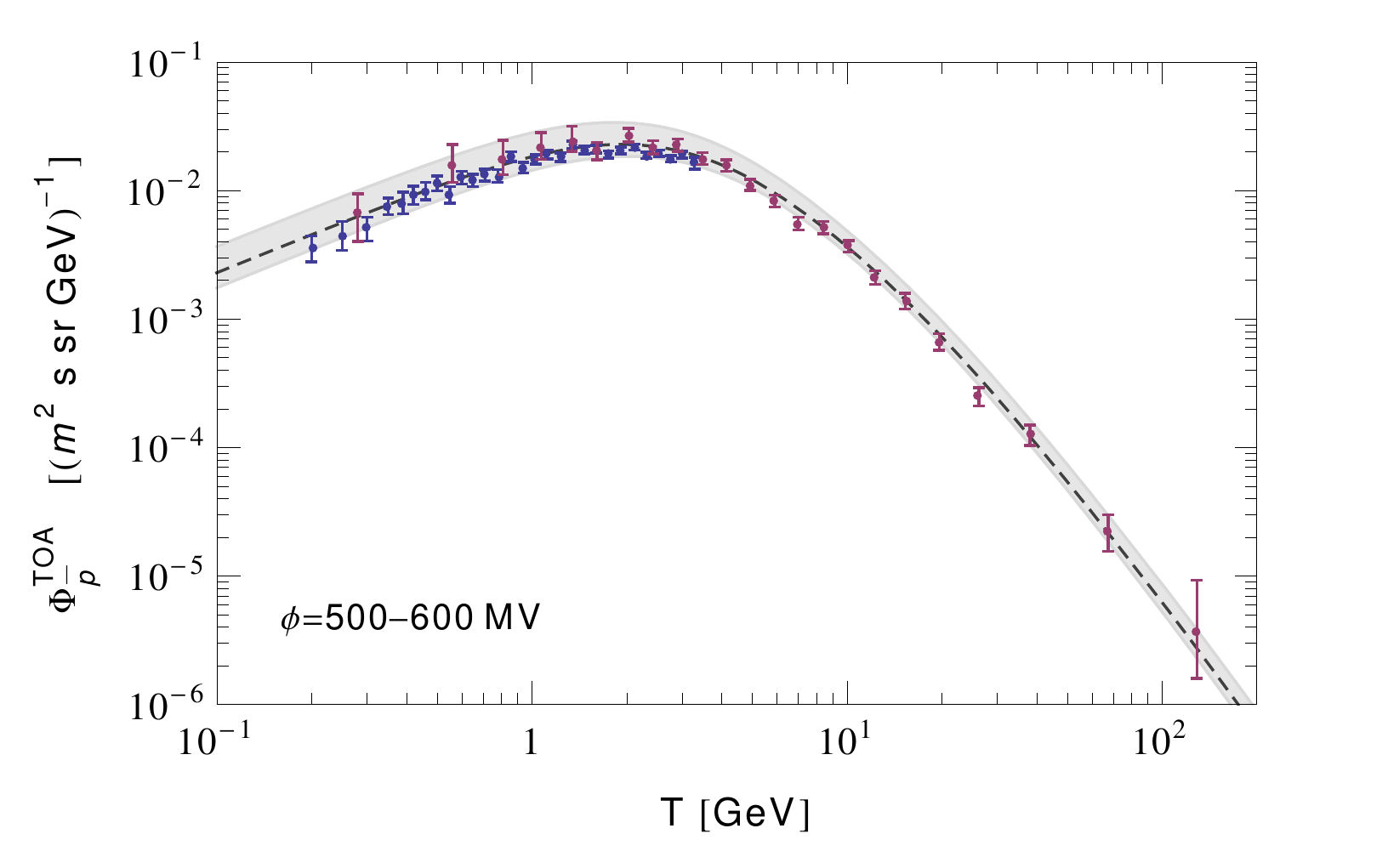}
\caption{Interstellar antiproton flux and particle physics uncertainty band obtained in this work (upper panel, also available in the Ancillary Files). In the lower panel, the top-of-the-atmosphere flux obtained by modulating $\Phi_{\bar{p}}^\text{IS}$ with a force field $\phi=500-600\:\text{MV}$ is compared with the experimental data from PAMELA and BESS-Polar II.}
\label{fig:finalflux}
\end{figure}

In figure~\ref{fig:finalflux} (upper panel), we depict the interstellar antiproton flux obtained from the source term of figure~\ref{fig:na49vstan} within the two-zone diffusion model. The shown error band comprises the particle physics uncertainties and the uncertainties of the primary cosmic ray fluxes, propagation uncertainties have not been included. The error band is practically inherited from the source term. Only at low $T$, the uncertainty in $\Phi_{\bar{p}}^\text{IS}$ is slightly reduced compared to the uncertainty in $q_{\bar{p}}$ due to the tertiary component of the flux.

To compare the obtained flux with existing experimental data, we have to account for solar modulation. For this we use the force field approximation. The currently most accurate measurements of the antiproton flux were performed by the PAMELA~\cite{Adriani:2010rc} and the BESS-Polar II~\cite{Abe:2011nx} collaborations. Both data sets were obtained in a period of low solar activity, where a force field of $\phi=500-600\:\text{MV}$ can be used to account for modulation~\cite{Potgieter:2013cwj}. In figure~\ref{fig:finalflux} (lower panel) we compare the top-of-the-atmosphere flux from our calculation with the experimental data of PAMELA and BESS-Polar II. It can be seen that a very good agreement within the given uncertainties is obtained. We notice that the measured flux resides more towards the lower end of the uncertainty band which seems to speak against strong isospin effects in the antineutron production. However, this conclusion might be premature as we have not included the propagation uncertainties in the antiproton flux.

\section{Conclusion}
The AMS-02 experiment is about to release a high precision measurement of the cosmic ray antiproton flux. In this light, we have rederived the secondary antiproton source term which arises from cosmic ray spallations in the galactic disc (figure~\ref{fig:na49vstan}). We employed new experimental data on proton proton and proton nucleus scattering recorded by the NA49 experiment at CERN. Important improvements of our study include the careful treatment of antiprotons arising from antineutron and hyperon decay as well as a revised approach to proton helium scattering. For the first time, we were able to assign a realistic particle physics error band to the secondary source term, which is an important input for constraining possible primary antiproton signals in the galaxy. Our source term is publically available and can be used with a generic propagation setup. 

We have identified two major sources of uncertainty in the secondary source term: one stems from the lack of experimental data on low energy proton collisions, where hadronic cross sections do not follow a scaling behavior. The other is related to a possible isospin effect which may result in an enhanced $\bar{n}$-production in $pp$ collisions. In this regard, the experimental situation, however, remains dubious and further investigation is required.

Independently, we determined the source term by use of the Monte Carlo tools PYTHIA, GEANT and DPMJET. While there appears a qualitative agreement with the data-based evaluation of the source term at antiproton energies $T>10\gev$, the Monte Carlo generators substantially overestimate $q_{\bar{p}}$. We have identified the origin of the discrepancy in the too large antiproton multiplicities predicted by the Monte Carlo generators. At $T\ll10\gev$ the predictions of all three Monte Carlo tools become rather poor which is most likely explained by the breakdown of the underlying hadronization models.

In the last step, we have propagated the source term within the two-zone diffusion framework in order to obtain the secondary antiproton flux. The latter was then confronted with existing experimental data by the PAMELA and BESS-Polar II experiments and a very satisfying agreement was obtained (figure~\ref{fig:finalflux}). The comparison with the upcoming AMS-02 data is eagerly awaited.

\section*{Note added}
In the final stage of this work we became aware of a related study by F. Donato, A. Goudelis, M. Di Mauro and P. Serpico~\cite{DGMS}. While it also determines the proton scattering cross sections relevant for cosmic ray physics, the approach is quite complementary to ours. We focus on the NA49 data sets which we use as a basis to discuss prompt and hyperon-induced antiproton production, antineutron production and proton nucleus scattering. The focus of~\cite{DGMS} on the other hand is on a global reanalysis of the existing $pp$ scattering data. In this light~\cite{DGMS}, provides a very important comparison for our work. 

\section*{Acknowledgments}
This work has been supported by the German Science
Foundation (DFG) within the Collaborative Research
Center 676 "Particles, Strings and the Early Universe" 
and the SFB-Transregio TR33 "The Dark Universe".

\bibliography{literature}

\providecommand{\bysame}{\leavevmode\hbox to3em{\hrulefill}\thinspace}
\begin{thebibliography}{10}

\bibitem{Aguilar:2013qda}
AMS Collaboration, M.~Aguilar et~al., Phys.Rev.Lett. \textbf{110} (2013),
  141102.

\bibitem{Tan:1982nc}
L.~Tan and L.~Ng, Phys.Rev. \textbf{D26} (1982), 1179--1182.

\bibitem{Duperray:2003bd}
R.~Duperray, C.-Y. Huang, K.~Protasov, and M.~Buenerd, Phys.Rev. \textbf{D68}
  (2003), 094017,  [astro-ph/0305274].

\bibitem{Anticic:2009wd}
NA49 Collaboration, T.~Anticic et~al., Eur.Phys.J. \textbf{C65} (2010), 9--63,
  [0904.2708].

\bibitem{Baatar:2012fua}
NA49 Collaboration, B.~Baatar et~al., Eur.Phys.J. \textbf{C73} (2013), 2364,
  [1207.6520].

\bibitem{Fischer:2003xh}
NA49 Collaboration, H.~Fischer, Heavy Ion Phys. \textbf{17} (2003), 369--386.

\bibitem{Chvala:2003dn}
NA49 Collaboration, O.~Chvala, Eur.Phys.J. \textbf{C33} (2004), S615--S617,
  [hep-ex/0405053].

\bibitem{Maurin:2001sj}
D.~Maurin, F.~Donato, R.~Taillet, and P.~Salati, Astrophys.J. \textbf{555}
  (2001), 585--596,  [astro-ph/0101231].

\bibitem{Donato:2001ms}
F.~Donato, D.~Maurin, P.~Salati, A.~Barrau, G.~Boudoul, et~al., Astrophys.J.
  \textbf{563} (2001), 172--184,  [astro-ph/0103150].

\bibitem{Simon:1998}
M.~Simon, A.~Molnar, and S.~Roesler, Astrophys.J. \textbf{499} (1998), 250.

\bibitem{Strong:1998pw}
A.~Strong and I.~Moskalenko, Astrophys.J. \textbf{509} (1998), 212--228,
  [astro-ph/9807150].

\bibitem{Sjostrand:2007gs}
T.~Sjostrand, S.~Mrenna, and P.~Z. Skands, Comput.Phys.Commun. \textbf{178}
  (2008), 852--867,  [0710.3820].

\bibitem{Roesler:2000he}
S.~Roesler, R.~Engel, and J.~Ranft,  (2000), 1033--1038,  hep-ph/0012252.

\bibitem{Agostinelli:2002hh}
GEANT4, S.~Agostinelli et~al., Nucl.Instrum.Meth. \textbf{A506} (2003),
  250--303.

\bibitem{Brun:1997pa}
R.~Brun and F.~Rademakers, Nucl.Instrum.Meth. \textbf{A389} (1997), 81--86.

\bibitem{Carey:1974gf}
D.~C. Carey, J.~Johnson, R.~Kammerud, M.~Peters, D.~Ritchie, et~al.,
  Phys.Rev.Lett. \textbf{33} (1974), 327--330.

\bibitem{Taylor:1975tm}
F.~Taylor, D.~C. Carey, J.~Johnson, R.~Kammerud, D.~Ritchie, et~al., Phys.Rev.
  \textbf{D14} (1976), 1217.

\bibitem{Feynman:1969ej}
R.~P. Feynman, Phys.Rev.Lett. \textbf{23} (1969), 1415--1417.

\bibitem{Low:1975sv}
F.~Low, Phys.Rev. \textbf{D12} (1975), 163--173.

\bibitem{Nussinov:1975mw}
S.~Nussinov, Phys.Rev.Lett. \textbf{34} (1975), 1286--1289.

\bibitem{Brodsky:1976mg}
S.~J. Brodsky and J.~Gunion, Phys.Rev.Lett. \textbf{37} (1976), 402--405.

\bibitem{Brodsky:1977bu}
S.~J. Brodsky and J.~Gunion, Phys.Rev. \textbf{D17} (1978), 848--857.

\bibitem{Alt:2005zq}
NA49 Collaboration, C.~Alt et~al., Eur.Phys.J. \textbf{C45} (2006), 343--381,
  [hep-ex/0510009].

\bibitem{Beringer:1900zz}
Particle Data Group, J.~Beringer et~al., Phys.Rev. \textbf{D86} (2012), 010001.

\bibitem{Barr:2006fs}
G.~Barr, O.~Chvala, H.~Fischer, M.~Kreps, M.~Makariev, et~al., Eur.Phys.J.
  \textbf{C49} (2007), 919--945,  [hep-ex/0606029].

\bibitem{Aaij:2012ut}
LHCb Collaboration, R.~Aaij et~al., Eur.Phys.J. \textbf{C72} (2012), 2168,
  [1206.5160].

\bibitem{Capiluppi:1974rt}
P.~Capiluppi, G.~Giacomelli, A.~Rossi, G.~Vannini, A.~Bertin, et~al.,
  Nucl.Phys. \textbf{B79} (1974), 189.

\bibitem{Alper:1975jm}
British-Scandinavian Collaboration, B.~Alper et~al., Nucl.Phys. \textbf{B100}
  (1975), 237.

\bibitem{Arsene:2007jd}
BRAHMS Collaboration, I.~Arsene et~al., Phys.Rev.Lett. \textbf{98} (2007),
  252001,  [hep-ex/0701041].

\bibitem{1900hyj}
J.~Smith,  (1972).

\bibitem{Allaby:1970jt}
J.~Allaby, F.~Binon, A.~Diddens, P.~Duteil, A.~Klovning, et~al.,  (1970).

\bibitem{Kalinovskii:1989}
A.~Kalinovskii, M.~Mokhov, and Y.~Nikitin, American Institute of Physics
  (1989).

\bibitem{Cronin:1973fd}
J.~Cronin, H.~J. Frisch, M.~Shochet, J.~Boymond, P.~Piroue, et~al.,
  Phys.Rev.Lett. \textbf{31} (1973), 1426--1429.

\bibitem{Letaw:1983}
J.~Letaw, R.~Silberberg, and C.~Tsao, Astrophys.J.S. \textbf{51} (1983), 271.

\bibitem{Putze:2010zn}
A.~Putze, L.~Derome, and D.~Maurin, Astron.Astrophys. \textbf{516} (2010), A66,
   [1001.0551].

\bibitem{Potgieter:2013cwj}
M.~Potgieter, E.~Vos, M.~Boezio, N.~De~Simone, V.~Di~Felice, et~al.,  (2013),
  1302.1284.

\bibitem{Shikaze:2006je}
Y.~Shikaze, S.~Haino, K.~Abe, H.~Fuke, T.~Hams, et~al., Astropart.Phys.
  \textbf{28} (2007), 154--167,  [astro-ph/0611388].

\bibitem{Ams:2013}
S.~Haino, \emph{{Talk at the 33rd {ICRC} Conference}}, 2013.

\bibitem{Gleeson:1968zza}
L.~Gleeson and W.~Axford, Astrophys.J. \textbf{154} (1968), 1011.

\bibitem{Antinucci:1972ib}
M.~Antinucci, A.~Bertin, P.~Capiluppi, M.~D'Agostino-Bruno, A.~Rossi, et~al.,
  Lett.Nuovo Cim. \textbf{6} (1973), 121--128.

\bibitem{Maurin:2002ua}
D.~Maurin, R.~Taillet, F.~Donato, P.~Salati, A.~Barrau, et~al.,  (2002),
  astro-ph/0212111.

\bibitem{Maurin:2002hw}
D.~Maurin, R.~Taillet, and F.~Donato, Astron.Astrophys. \textbf{394} (2002),
  1039--1056,  [astro-ph/0206286].

\bibitem{Kappl:2011jw}
R.~Kappl and M.~W. Winkler, Phys.Rev. \textbf{D85} (2012), 123522,
  [1110.4376].

\bibitem{Moskalenko:2001ya}
I.~V. Moskalenko, A.~W. Strong, J.~F. Ormes, and M.~S. Potgieter, Astrophys.J.
  \textbf{565} (2002), 280--296,  [astro-ph/0106567].

\bibitem{Evoli:2008dv}
C.~Evoli, D.~Gaggero, D.~Grasso, and L.~Maccione, JCAP \textbf{0810} (2008),
  018,  [0807.4730].

\bibitem{Lavalle:2014kca}
J.~Lavalle, D.~Maurin, and A.~Putze,  (2014),  1407.2540.

\bibitem{Bringmann:2011py}
T.~Bringmann, F.~Donato, and R.~A. Lineros, JCAP \textbf{1201} (2012), 049,
  [1106.4821].

\bibitem{Lavalle:2010yw}
J.~Lavalle, Phys.Rev. \textbf{D82} (2010), 081302,  [1007.5253].

\bibitem{Evoli:2011id}
C.~Evoli, I.~Cholis, D.~Grasso, L.~Maccione, and P.~Ullio, Phys.Rev.
  \textbf{D85} (2012), 123511,  [1108.0664].

\bibitem{Bringmann:2006im}
T.~Bringmann and P.~Salati, Phys.Rev. \textbf{D75} (2007), 083006,
  [astro-ph/0612514].

\bibitem{Adriani:2010rc}
PAMELA Collaboration, O.~Adriani et~al., Phys.Rev.Lett. \textbf{105} (2010),
  121101,  [1007.0821].

\bibitem{Abe:2011nx}
K.~Abe, H.~Fuke, S.~Haino, T.~Hams, M.~Hasegawa, et~al., Phys.Rev.Lett.
  \textbf{108} (2012), 051102,  [1107.6000].

\bibitem{DGMS}
F.~Donato, A.~Goudelis, M.~Di~Mauro, and P.~Serpico, to appear.

\end{thebibliography}
\bibliographystyle{ArXiv}
\end{document}